\documentclass[a4paper,11pt]{article}
\usepackage[dvipdfmx]{graphicx}
\usepackage{graphicx}
\usepackage{dcolumn}
\usepackage{bm}
\usepackage{multirow}
\usepackage{color}
\usepackage{url}
\usepackage{xcolor}
\usepackage[italic]{hepnames}
\usepackage{tabularx}
\usepackage{array,booktabs}
\usepackage{caption}
\usepackage{subcaption}
\graphicspath{{figs/}}  
\usepackage{adjustbox}
\usepackage{placeins}
\usepackage{lineno}
\usepackage{enumitem}
\usepackage[subpreambles=true]{standalone}
\usepackage{gensymb}
\usepackage{float}
\usepackage{hhline}
\usepackage{jinstpub}
\usepackage{hyperref}

\newcommand{\nai}{NaI(Tl)}

\newcommand{\supltemp}{22$\degree$C}

\newcommand{\kevee}{keV$_{\text{ee}}$}
\newcommand{\Ba}{$^{133}$Ba}

\newcommand{\DSA}[1]{\textcolor{black}{BC0174}}
\newcommand{\DSB}[1]{\textcolor{black}{BC0175}}
\newcommand{\DSC}[1]{\textcolor{black}{combined}}

\newcommand{\MDA}[1]{\textcolor{black}{BC0174}}
\newcommand{\MDB}[1]{\textcolor{black}{BC0175}}
\newcommand{\MDC}[1]{\textcolor{black}{Combined}}

\definecolor{mj-oj}{RGB}{209, 94, 6}

\title{Characterisation of Hamamatsu R11065-20 PMTs for use in the SABRE South NaI(Tl) Crystal Detectors}

\abstract{The SABRE Experiment is a direct detection dark matter experiment using a target composed of multiple NaI(Tl) crystals. The experiment aims to be an independent check of the DAMA/LIBRA results with a detector in the Northern (Laboratori Nazionali Del Gran Sasso, LNGS) and Southern (Stawell Underground Physics Laboratory, SUPL) hemispheres. 
The SABRE South photomultiplier tubes (PMTs) will be used near the low energy noise threshold and require a detailed calibration of their performance and contributions to the background in the \nai{} dark matter search, prior to installation. We present the development of the pre-calibration procedures for the R11065-20 Hamamatsu PMTs. These PMTs are directly coupled to the NaI(Tl) crystals within the SABRE South experiment. In this paper we present methodologies to characterise the gain, dark rate, and timing properties of the PMTs. We develop a method for in-situ calibration without a light injection source. Additionally we explore the application of machine learning techniques using a Boosted Decision Tree (BDT) trained on the response of single PMTs to understand the information available for background rejection. Finally, we briefly present the simulation tool used to generate digitised PMT data from optical Monte Carlo simulations.
   }

\affiliation[1]{School of Physics, The University of Melbourne, Parkville, VIC 3010, Australia}
\affiliation[2]{Department of Nuclear Physics and Accelerator Applications, The Australian National University, Canberra, ACT 2601, Australia}
\affiliation[3]{Department of Physics, The University of Adelaide, Adelaide, SA 5005, Australia}
\affiliation[4]{Centre for Astrophysics and Supercomputing, Swinburne University of Technology, Hawthorn, VIC 3122, Australia}
\affiliation[5]{School of Physics, The University of Sydney, NSW 2006 Camperdown, Sydney, Australia}
\affiliation[6]{ARC Centre of Excellence for Dark Matter Particle Physics, Australia}
\affiliation[7]{International Center for Quantum-field Measurement System for Studies of the Universe and Particles (QUP), High-Energy Accelerator Research Organization (KEK), Oho, Tsukuba, Ibaraki 305-0801, Japan}
\affiliation[8]{INFN Sezione di Milano, via Celoria 16, 20133 Milano, Italy}
\affiliation[9]{SUBATECH, IMT Atlantique, CNRS/IN2P3, University of Nantes, 44307 Nantes, France}

\author[1,6,9,a]{O.~Stanley,}
\author[1,6]{W.~J.~D.~Melbourne,}
\author[1,6,b]{P.~Urquijo,}
\author[1,6]{E.~Barberio,}
\author[2,6]{V.~U.~Bashu,}
\author[2,6]{L.~J.~Bignell,}
\author[3,6,8]{I.~Bolognino,}
\author[4,6]{G.~Brooks,}
\author[1,6]{S.S.~Chhun,}
\author[2,6]{F.~Dastgiri,}
\author[2,6]{M.~B.~Froehlich,}
\author[5,6]{T.~Fruth,}
\author[1,6]{G.~Fu,}
\author[3,6]{G.~C.~Hill,}
\author[1,6]{R.~S.~James,}
\author[3,6]{K.~Janssens,}
\author[5,6]{S.~Kapoor,}
\author[2,6]{G.~J.~Lane,}
\author[3,6]{K.~T.~Leaver,}
\author[1,6]{J.~McKenzie,}
\author[2,6]{L.~J.~McKie,}
\author[3,6]{P.~McGee,}
\author[2,6]{P.~C.~McNamara,}
\author[1,6]{M.~Mews,}
\author[1,6]{L.~J.~Milligan,}
\author[1,6]{K.~J.~Rule,}
\author[4,6]{F.~Scutti,}
\author[2,6]{Z.~Slavkovsk\'{a},}
\author[2,6]{A.~E.~Stuchbery,}
\author[7]{B.~Suerfu,}
\author[1,6]{G.~N.~Taylor,}
\author[3,6]{A.~G.~Williams,}
\author[1,6]{Y.~Xing,}
\author[2,6]{Y.Y.~Zhong,}

\collaboration{SABRE South Collaboration}
\emailAdd{sabre-contact@lists.unimelb.edu.au}
\emailAdd{(a) ostanley@student.unimelb.edu.au}
\emailAdd{(b) purquijo@unimelb.edu.au}

\begin{document}
    \maketitle

\section{Introduction}
The SABRE (Sodium iodide with Active Background REjection)~\cite{SABRE_POP,SABRE_POPMC,Barberio_2025,SABRE_MC} South experiment is a \nai{} dark matter (DM) direct detection experiment designed to provide a model-independent test of the long-standing modulating signal observed by the DAMA/LIBRA experiment~\cite{DAMA2020_achievements_perspectives}, with sensitivity to confirm or reject this result. DAMA/LIBRA observes a modulation in the recoil rate in \nai{} in the 1-6~\kevee{} electron equivalent energy region, corresponding to about 5.5~--~45 photoelectrons (PE)~\cite{DAMA2020_achievements_perspectives} with DAMA/LIBRA's light yield (LY). The observation is consistent with an annual modulation due to DM interactions. This result is the only evidence for direct detection of DM, with current NaI(Tl) experiments being in tension with the DAMA/LIBRA observations~\cite{ANAIS_3_year_mod,COSINE_3_year_mod}. Other direct detection experiments have produced more stringent constraints on the spin-independent cross section for WIMP-like dark matter models, making use of different target materials and detector designs~\cite{Aalbers_2023, Aprile_2023}. Of those that have searched for annual modulations~\cite{XMASS:2018koa,DarkSide-50:2023fgf}, none have found a signal consistent with the DAMA/LIBRA observation. 

SABRE South will be located at the Stawell Underground Physics Laboratory (SUPL) with 1025~m of flat rock overburden providing 2870~m of water equivalent shielding~\cite{Barberio_2025}. The SABRE South experiment has, as DM target, a set of 7~$\times$~7~kg ultra-high purity NaI(Tl) crystals~\cite{SABRE_NAI33Characterisation,Barberio_2025}, each directly coupled to two 76~mm Hamamatsu Photonics (HPK) R11065-20 PMTs encased in enclosures made with high purity oxygen-free copper. The enclosures are flushed with nitrogen to prevent contamination from potential outgassing of Teflon components, other internal sources of $^{210}$Pb, and to protect the \nai{} crystals, which are highly soluble in water. The enclosure will be submerged within an active veto system with $11.6$~m$^3$ of linear-alkylbenzyene liquid scintillator~\cite{ABUSLEME2021164823} instrumented with at least 18~$\times$~204~mm HPK R5912 PMTs\footnote{Additional PMTs from the decommissioned Daya Bay experiment have been acquired and are being considered for installation.}. The vessel is encased in a passive shield comprised of steel and polyethylene layers which shields from external radiation and neutrons. A muon veto system comprised of eight EJ200 scintillators (40~$\times$~300~$\times$~5~cm) each coupled to two HPK R13089 PMTs covering a total area of 9.6~m$^2$ is placed on top of the passive shielding. The detector will operate at a temperature of \supltemp{}.

Currently running \nai{} experiments have observed rate excesses above the expected background models at energies in the DAMA region of interest (RoI) of 1~--~6~\kevee{}~\cite{ANAIS_3_year_mod,COSINE-100:2020wrv}. These excesses are most prominent in the 1~--~3~\kevee{} region and become small above 3~\kevee{}.  A potential origin of this background is noise introduced by the PMTs and the data acquisition readout system~\cite{Coarasa_2022}. This may also bias the annual modulation search, as the residual rate for some searches is determined by subtracting a time-dependent background model~\cite{James:2024gtl,COSINE_InducedModulations}. 

\begin{figure}[htb]
    \centering
    \includegraphics[height = 6cm]{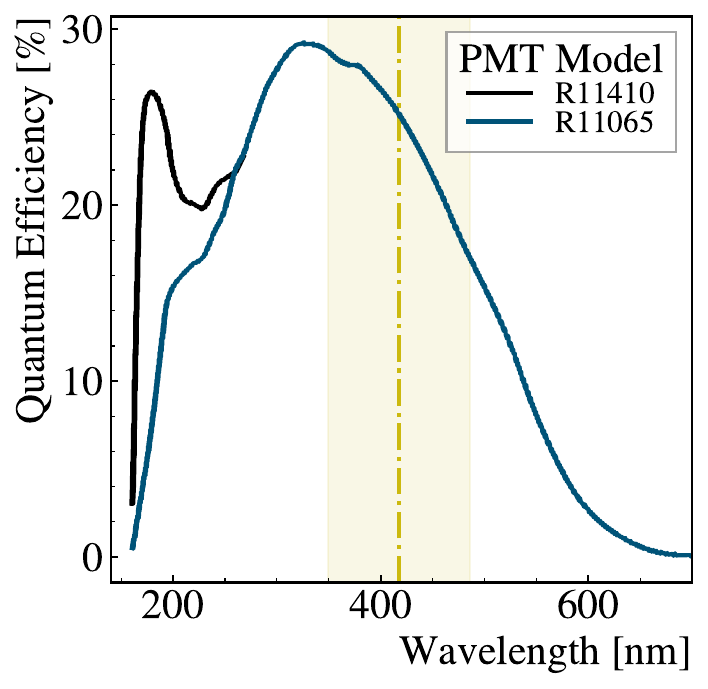}
    \caption{The typical spectral response for the R11065 and R11410 PMTs\cite{R_Acciarri_2012}. The shaded region shows where the emission spectrum of NaI(Tl) is above 50$\%$ of the peak emission amplitude. The vertical line indicates the emission peak at 415 nm.}~\cite{HamSpecSheet}
    
    \label{fig:bialkali-emission}
\end{figure}

To be sensitive to a signal in the DAMA/LIBRA RoI (1~--~6 \kevee{}), PMTs are required to operate stably with sensitivity to single photons throughout the lifetime of the experiment, without making a significant contribution to the experimental background. The HPK R11065 PMTs were selected for the SABRE South experiment due to their low dark rate, high quantum efficiency (QE)~\cite{HamSpecSheet} (the typical \nai{} LY, for a PMT with 30~$\%$ QE is 12~PE~keV$^{-1}$) and high efficiency around the emission spectrum for \nai{} shown in Fig.~\ref{fig:bialkali-emission}. 

The HPK R11065 PMT was developed together with the HPK R11410 as both PMTs are designed for use in low background experiments. The HPK R11410 was optimised for very ultra-violet (VUV) emissions in liquid xenon and operated by dark matter detectors such as XENONnT and LUX-ZEPLIN (LZ)~\cite{XENON:2015ara,Aalbers_2023}. The HPK R11065 was developed for operation with liquid argon using a wavelength shifter to 420 nm and is used in the DarkSide experiment~\cite{R_Acciarri_2012,DarkSide-50:2023fgf}.  

Each SABRE PMT was selected by the manufacturer to have a QE~$>30\%$. Accurate characterisation of the HPK R11065-20 PMT's gain, dark rate, timing, and after-pulsing properties are crucial to help understand systematic errors in event selection and energy scale, the degradation in performance of the detectors over time, and to inform methods for long term in-situ calibration. Pre-calibration allows for control of the experimental conditions by making use of reference detectors (silicon photomultipliers (SiPMs) and scintillator panels) as well as a controlled source (pico second pulsed laser). Several pre-calibration measurements (such as transit-time) cannot be absolutely calibrated in-situ (or will be dominated by other signals), and others such as gain can be calibrated directly with a light source rather than via scintillation. In contrast to the PMTs of the SABRE South veto system, the crystal system will not have any in-situ optical calibration system.

In this paper, we describe the methods and procedures that will be applied in the pre-calibration of the SABRE South R11065 PMTs. This is demonstrated on two of the PMTs with serial numbers BC0174 and BC0175 (delivered in 2020 from HPK). All PMTs purchased by SABRE have been tested by HPK to have QEs between 30.06~$\%$ and 33.25~$\%$ at 420 nm. The pre-calibration measurements are used in SABRE South Monte Carlo detector simulations to understand the PMT related background contribution in the RoI, and to determine the performance of background suppression classifiers. We also present the current method developed to discriminate between signal (scintillation) and background (dark count) in the RoI, using information from single PMTs (without requiring any coincidence information). We additionally test a BDT using both PMTs simultaneously. Both methods uses variables that has been explored by previous \nai{} analyses~\cite{Coarasa_2022, SABRE_NAI33Characterisation, Spinks_2023}. Finally, we briefly describe the waveform simulation tool developed for the SABRE South simulation software framework.

\section{Experimental Setup} \label{sec:exp_setup}
 The R11065-20 is a 76~mm, box and line 12 dynode stage, head-on PMT with a mu-metal alloy body. The PMT has a typical transit-time of 46 ns with a typical gain of 5~$\times$~10$^{6}$ at 1500~V~\cite{HamSpecSheet}. The PMTs have photocathodes chosen to operate with a peak response at 420~nm, which is well matched to the peak emission of \nai{} (415~nm).  
 
\begin{figure}[htb]
        \centering
        \captionsetup[subfigure]{justification=centering}
        \includegraphics[width=0.32\columnwidth]{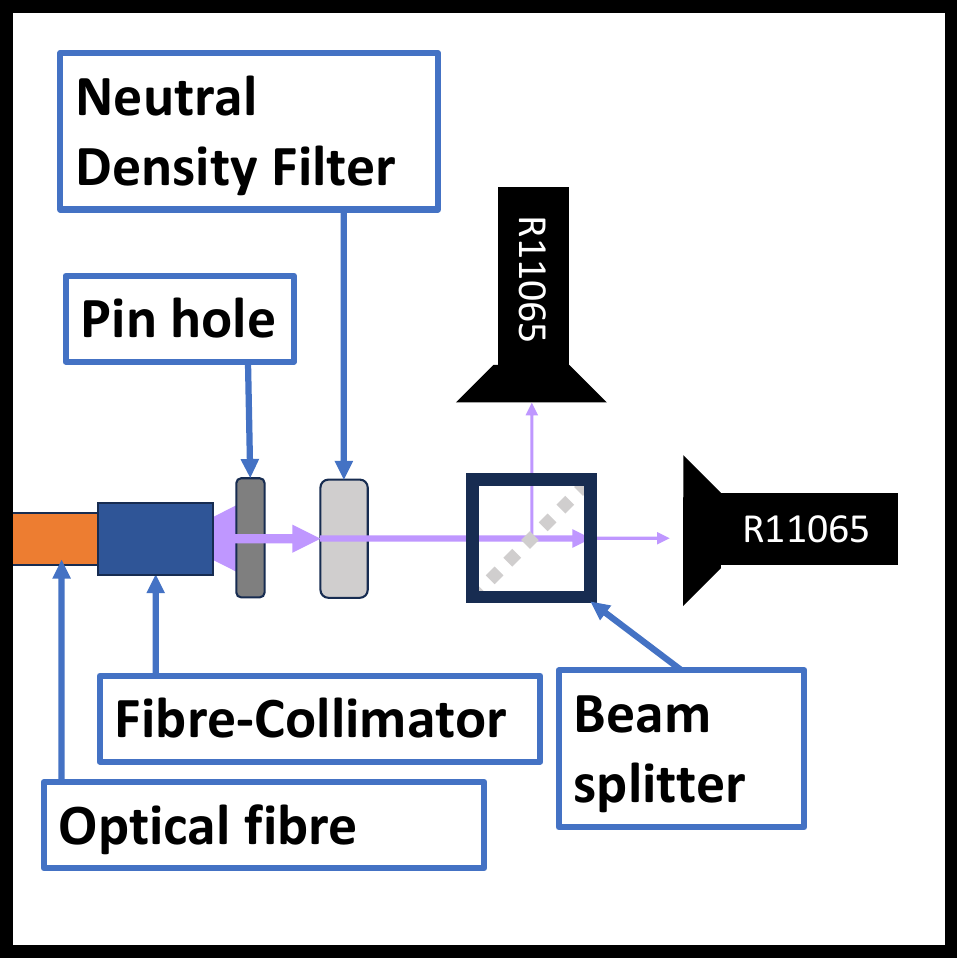}
        \hfill
             \includegraphics[width=0.32\columnwidth]{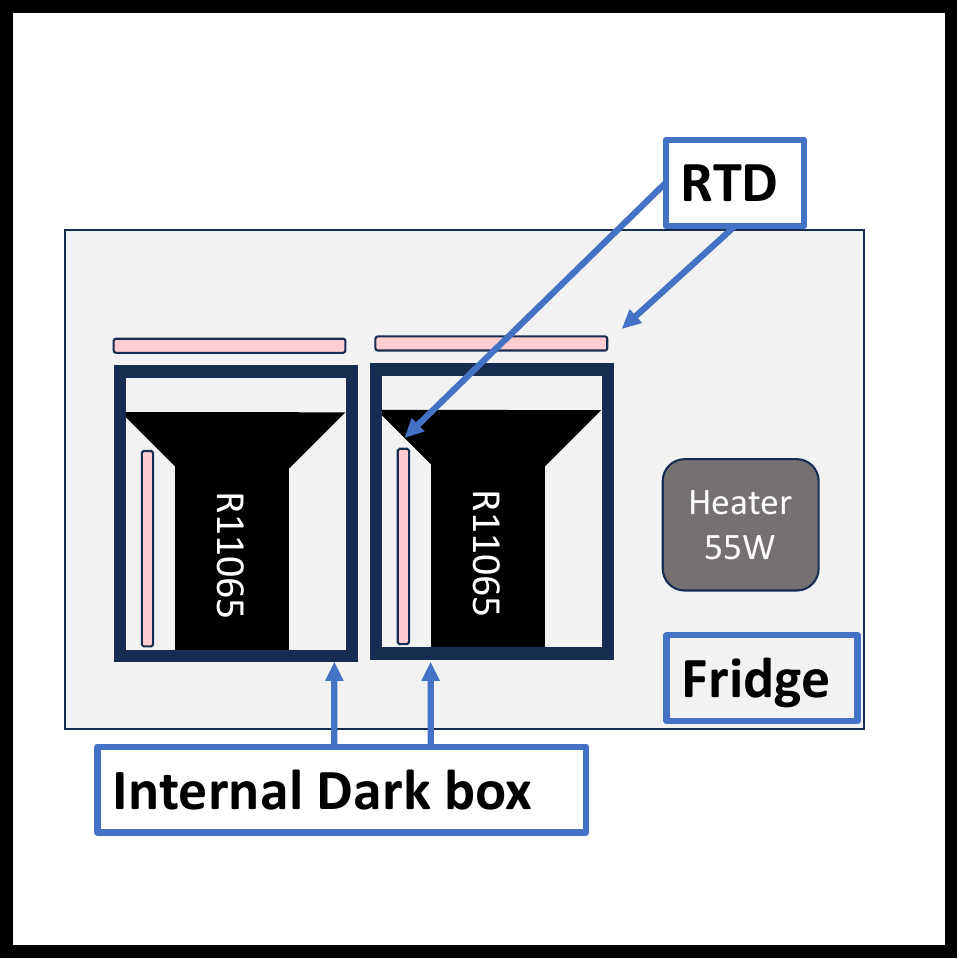}
        \hfill
             \includegraphics[width=0.32\columnwidth]{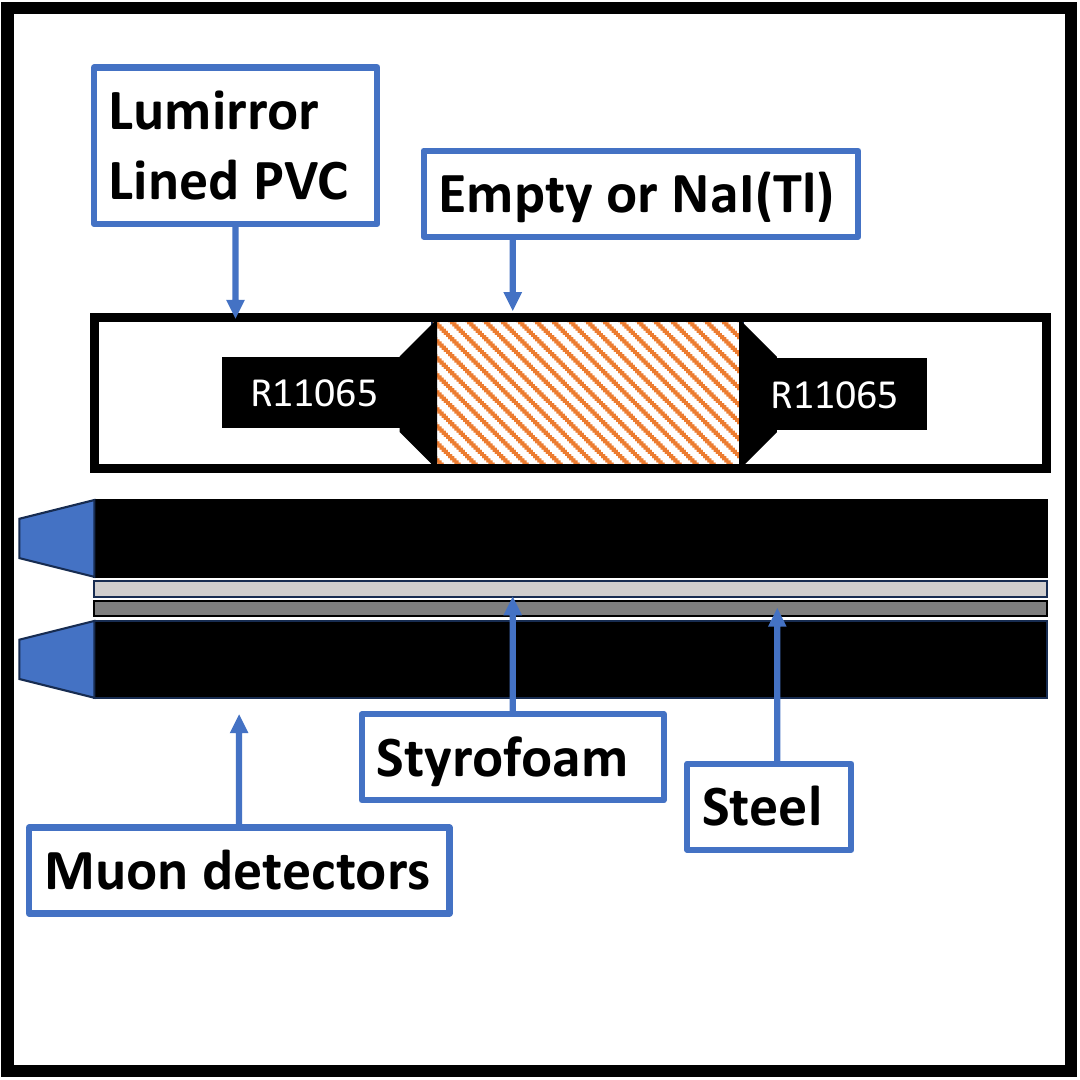}
        \caption{Experimental set ups for (left) pico second pulsed laser system, (middle) thermal testing chamber with resistance temperature device sensors (right) SABRE detector mockup, operated with and without an encapsulated commercial NaI(Tl) crystal. }
        \label{fig:ExpSetup}
    \end{figure}

 The PMT characterisation measurements discussed in this paper are performed using three test systems at the University of Melbourne: (i) a pico second pulsed laser system, (ii) a thermal testing chamber, and (iii) a SABRE South NaI(Tl) detector mockup. The laser system, shown in Fig.~\ref{fig:ExpSetup} (left), is housed in an aluminium dark box fixed to an optical table that acts as a Faraday cage, with feedthroughs for signal and HV connections, and a multimode optical fibre with a fibre collimator.  An HPK PLP-10 pulsed laser (405 nm)~\cite{HamLaserSpecSheet} light source with a pulse width of 60~ps is injected through the optical fibre and attenuated by a neutral density (ND) filter (10$^{-4}$). The PMTs are optically shielded with non-reflective matte black material. For timing and after-pulse measurements, an SiPM (HPK S10362-11-050) replaces the perpendicular PMT, to be used as a reference detector. The SiPM has an effective area of 1~mm$^2$  with transit-time resolution (FWHM)  of 200 - 300 ps as specified by HPK. 

The thermal testing system (Fig.~\ref{fig:ExpSetup} (middle)) consists of a chest freezer with a 55~W heater and an electronically controlled thermostat to control the two devices. The PMT is optically shielded in a small dark box within the thermal chamber. Four resistance temperature device (RTD) sensors (PT100 IEC 60751 class 1/10) are used in the thermal chamber. The error on the RTD is described by
\begin{equation} \label{eq:RTD_T_Err}
    \delta T = 0.03 + 0.005 \times T.
\end{equation}
where $T$ is the measured temperature in~$\degree$C (i.e., at 22~$\degree$C the uncertainty is 0.14~$\degree$C)~\cite{RTD_Krishnaan}. Each PMT has one sensor in contact with the external Teflon wrapping on the stem of the PMT, and two more sensors are used to monitor the temperature within the freezer, close to and far from the heating element, to provide information on the distribution of temperature in the freezer. This system is operated between 15~--~30~$\degree$C, which is well beyond the expected operating range of SABRE South of $23\pm2$~$\degree$C. Before starting measurements, the PMTs were left in the dark at 1300~V for 12~hours to stabilise. At each voltage step they were left to stabilise for one hour. At each change in temperature (starting from 15~$\degree$C) a minimum of 3 hours was allowed for the PMTs to reach equilibrium with the thermal chamber. Each run is taken over a period of 100~s. All data were taken with the heating and cooling elements turned off, to ensure that heating and cooling do not occur during the data taking window, as these systems were found to introduce background electronic noise.  

 The SABRE NaI(Tl) detector mock-up enclosure (Fig.~\ref{fig:ExpSetup} (right)) is positioned within an aluminium dark box. It comprises a polyvinyl-chloride (PVC) tube lined with a Lumirror reflector and a pair of HPK R11065 PMTs. This system is used for the study of coincident signals between pairs of PMTs. The two PMTs are placed 100~mm apart: this gap is left empty for the study of dark coincident effects. We use a commercial NaI(Tl) crystal (an aluminium and quartz encapsulated 75~mm diameter, 100~mm long crystal from EPIC crystals) to obtain crystal scintillation events. The crystal is not directly coupled to the PMTs, and instead a 1 mm air gap is maintained using a 3D printed mount for easily repeatable testing. 
 
 The SABRE enclosure mock-up is placed on top of two plastic scintillators (60~$\times$30~$\times$5~cm), with each coupled to a single Hamamatsu R7724 PMT. A 2.5~cm layer of polystyrene (which increases the gap between the scintillators) and a 1.5~cm layer of steel are placed between the scintillators. This allows for coincident muon tagging with high efficiency. Muons passing through the crystal and/or the PMT (window, vacuum, dynode, etc.) can result in the emission of light/electron in the crystal or PMT respectively. This is a significant portion of the data, as the system is located above ground. Tagging these events allows us to separate out muon events, to better replicate conditions underground. 

The nominal data acquisition (DAQ) systems of the SABRE South (and North) NaI(Tl) detector use 8 channel CAEN V1730D digitisers with a sampling rate of $500~\rm{MS}/s$ and $0.12~\mathrm{mV}$ resolution. The digitiser can trigger on individual PMTs, coincidences within a single module, and coincidences between modules. These triggers in turn can be used to form coincidence triggers with other systems, such as the liquid scintillator veto, to tag intrinsic radioactivity in the crystal. 

The sampling rate and resolution of the digitisers provide good dynamic range, and they can support recording of individual waveforms. This is crucial for signal-background separation near the RoI. The digitisers are connected to a high-performance computer via a CONET optical fibre connection and a CAEN A3818 PCIe card. Data collection is managed by custom DAQ software  written for the SABRE South experiment, which records raw waveforms~\cite{Barberio_2025} to be processed with the Pyrate framework~\cite{Scutti_2023}. The PMT high voltage is supplied using a CAEN SY5527 universal multichannel power supply mainframe system with A7435 power supplies, which can supply up to 3.5~kV (the recommended safe limit 1750~V for the R11065) and 3.5~mA per channel and has a built-in EPICS input-output controller.  
\section{Signal}
\subsection{Single Photo-Electron Response and Gain} \label{sec:spe}

The signal amplification behaviour of PMTs is determined by their gain, defined as the multiplication factor between the number of photoelectrons emitted from the cathode and the number of electrons in the PMT signal output at the end of the dynode chain. This can be extracted from the charge spectra acquired by illuminating the PMTs with single photons. We measure the single photoelectron (SPE) response of the detector by using the pico second pulsed laser setup shown in Fig.~\ref{fig:ExpSetup} (left). The laser is operated at a frequency of 10~kHz leading to a detection rate of approximately 1~kHz. A mean photon occupancy of $<$~0.1 photons per pulse is achieved, ensuring that more than 90~$\%$ of the detected signals are single photons. The PMT waveform and a synchronisation signal from the laser unit are acquired using the DAQ system outlined in Sec.~\ref{sec:exp_setup} with a hardware trigger threshold on the PMT waveform of 1.5~mV above the baseline corresponding to 30~$\%$ of the mean peak height of an SPE pulse at 1500~V for BC0174. This provides datasets that consist of SPE pulses and dark noise in roughly equal measure.

    \begin{figure}[htb]
        \centering
        \includegraphics[height = 6cm]{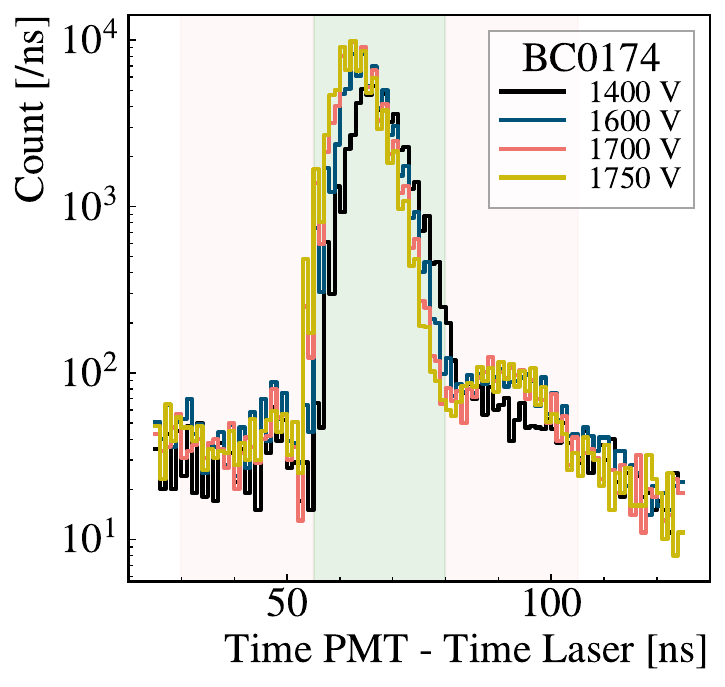}
        \includegraphics[height =6cm]{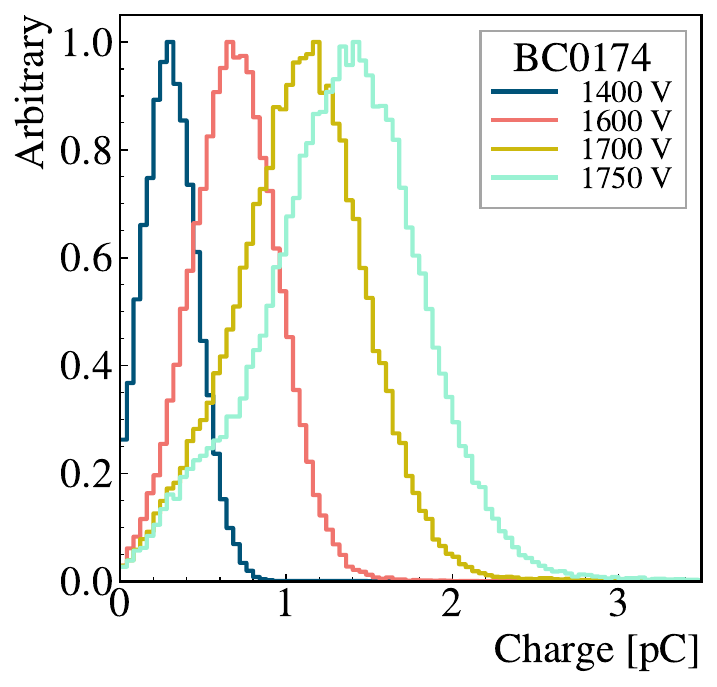}
        \caption{(left) The time-difference  between the pulsed-laser signal and the peak start location within the waveform for PMT BC0174. The green region is the selected region (55-80~ns) for our SPE analysis, and the pink are our side band region to measure purity. (right) Single photo-electron charge observed for PMT BC0174, normalised to 1 at SPE peak.}
        \label{fig:timediff}
    \end{figure}

    \subsubsection*{Single photon data selection}

\begin{figure}[htb]
    \centering
    \includegraphics[height = 6cm ]{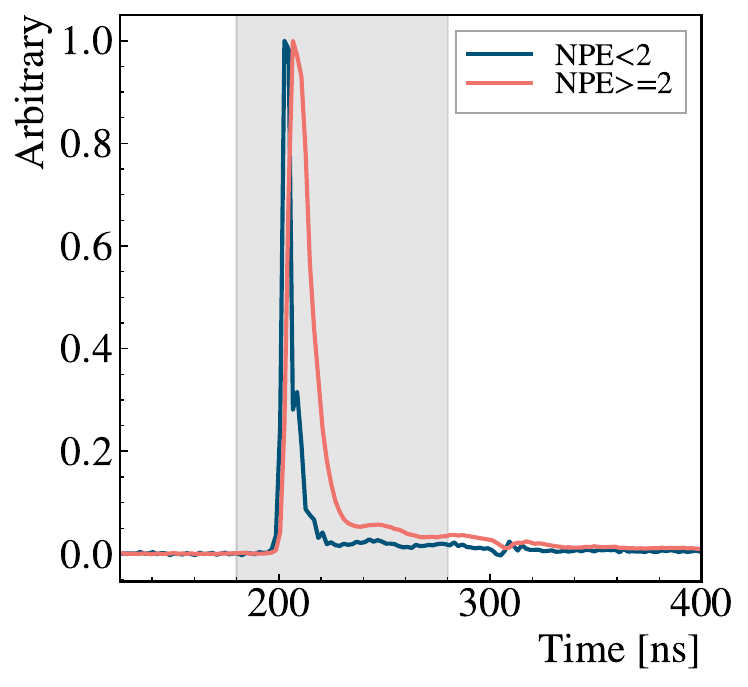}
    \includegraphics[height = 6cm ]{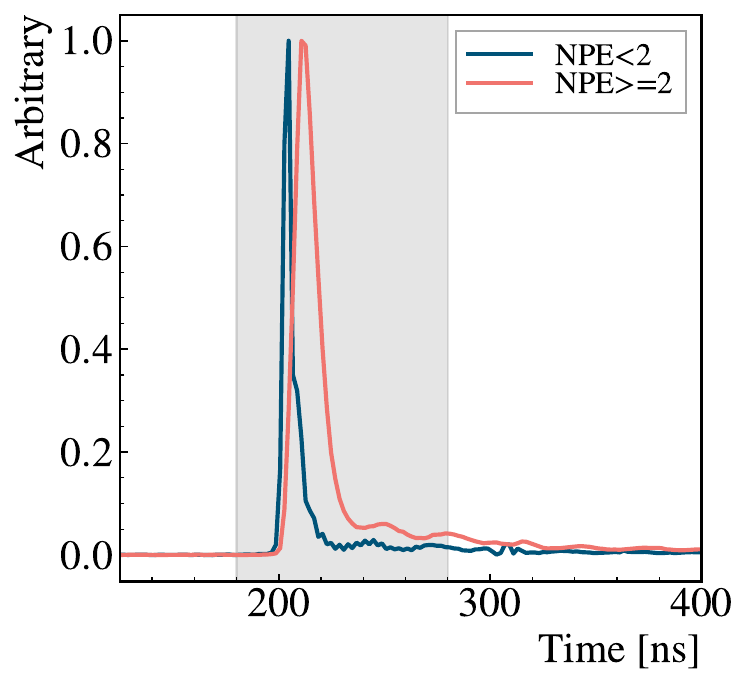}
    \caption{The normalised average waveforms for BC0174 (left) and BC0175 (right) at 1500~V, aligned at the leading edge threshold position (200 ns). The shaded region shows the charge integration window.}
    \label{fig:AverageWaveform_ChargeWindow}
\end{figure}

     To obtain a pure sample of laser-induced events, the requirement is that there is a coincident laser-synchronisation signal, identified based on peak height. The time difference between the PMT and the laser is shown in Fig.~\ref{fig:timediff} (left): we require it to be between 55~--~80 ns. The observed secondary peak is not understood and is of unknown origin. The spread in the time-difference distribution is a result of a number of different factors, including jitter in the laser synchronisation output ($\approx$ 10~ns), as well as the transit-time spread ($\approx$~9~ns see Sec.~\ref{sec:timing}).  Sample purity is estimated by counting events in two timing side bands (30~--~55 and 80~--~105 ns) normalised by window size, finding a signal purity > 94~$\%$ (Table~\ref{tab:r11065_gain}). The nonzero value of the time difference is largely attributed to the transit-time ($\approx$ 46 ns for R11065~\cite{kaplan2014hamamatsu} as measured in Sec.~\ref{sec:timing}), as well as the synchronous signal pulse width (5~$\pm$~3~ns, for the HPK PLP-10)~\cite{HamLaserSpecSheet} and a difference in cable length of 3~m between the PMT and laser synchronisation signals. Photo-electron charge is calculated for SPE events by integrating over the event waveform, from -20~--~+80~ns with respect to the identified pulse start location using a leading edge threshold of 1.5 mV. Figure~\ref{fig:AverageWaveform_ChargeWindow} shows the average waveforms in the SPE data set for events observed to have more or less than two PE. This extended range captures fluctuations in the tail of the distributions. It can result in a lower signal resolution, as we capture the fluctuations in the baseline. This was deemed sufficient to obtain a reliable measure of the SPE value. 

     The charge distributions after applying the selection criteria are shown in Fig.~\ref{fig:timediff}.  At higher voltages (above 1600~V) there is an additional contribution below the SPE peak due to underamplified signals (denoted ``uamp'')~\cite{PBCoates_1970}. This effect is only observed to make a significant contribution at higher voltages, as at lower voltages these events do not cross the trigger threshold.

    \subsubsection*{Analysis}
  
\begin{table}[htb]
    \centering
    \caption{SPE fit results, reporting the sample purity, charge mean ($\mu_{\rm SPE}$), standard deviation ($\sigma_{SPE}$), and the fraction of under-amplified events. The error quoted accounts for the fit and bias due to trigger thresholds. The error on the ratio of under amplified and SPE events are based on the fit errors. In the case where N$_{uamp}$  $\ll$ 1 we ignore this component. }
    \label{tab:r11065_gain}
    \renewcommand{\arraystretch}{1.1} 
    \begin{tabular}{|l|r|ccc|c|}
    \hline
    \textbf{PMT} & \textbf{Voltage} [V] & \textbf{Purity} [$\%$]& \textbf{$\mu_{\rm SPE}$} [pC/PE] & \textbf{$\sigma_{\rm SPE}$} [pC/PE] & \textbf{$N_{\rm uamp}/N_{\rm SPE}$} \\
    \hhline{|=|=|====|}
    \textbf{BC0174} & 1500 & 96.8  & 0.4405  $\pm$  0.0007 & 0.1936  $\pm$  0.0005 & - \\
     & 1550 & 97.3 & 0.5491  $\pm$  0.0008 & 0.2203  $\pm$  0.0006 & - \\
     & 1600 & 97.5  & 0.6941  $\pm$  0.0028 & 0.2565  $\pm$  0.0016 & 0.0158 $\pm$ 0.0059 \\
     & 1700 & 98.0  & 1.1104  $\pm$  0.0024 & 0.3538  $\pm$  0.0017 & 0.0555 $\pm$ 0.0070 \\
     & 1750 & 98.0  & 1.3772  $\pm$  0.0022 & 0.4218  $\pm$  0.0017 & 0.0610 $\pm$ 0.0056 \\
    \hhline{|=|=|====|}
    \textbf{BC0175} & 1500 & 94.0  & 0.5809  $\pm$  0.0012 & 0.2351  $\pm$  0.0009 & - \\
     & 1550 & 95.4  & 0.7465  $\pm$  0.0024 & 0.2698  $\pm$  0.0017 & 0.0140 $\pm$ 0.0072 \\
     & 1600 & 96.0  & 0.9671  $\pm$  0.0037 & 0.3142  $\pm$  0.0024 & 0.0533 $\pm$ 0.0140 \\
     & 1700 & 97.0  & 1.5399  $\pm$  0.0032 & 0.4508  $\pm$  0.0025 & 0.0668 $\pm$ 0.0090 \\
     & 1750 & 96.7  & 1.9049  $\pm$  0.0036 & 0.5501  $\pm$  0.0030 & 0.0792 $\pm$ 0.0135 \\
    \hline
    \end{tabular}
\end{table}

\begin{figure}[htb]
        \captionsetup[subfigure]{justification=centering}
        \centering
        \includegraphics[height=6 cm]{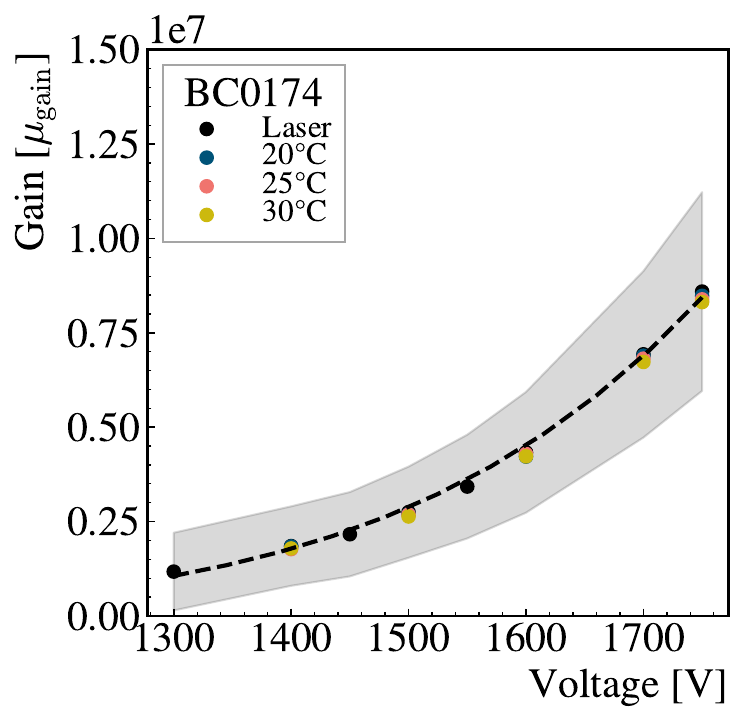}
        \includegraphics[height = 6cm]{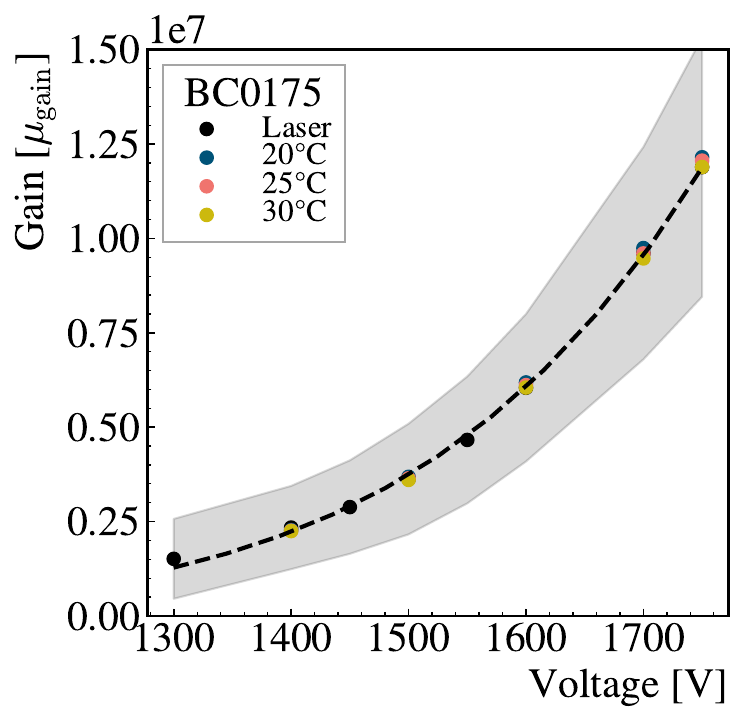}
        
        \caption{Gain values ($\mu_{\rm gain}$) as a function of operating voltage and data taking conditions. The dashed line displays the fit to the laser (Equation~\ref{eq:gainfit} ) calibrated data set taken at 20~$\degree$C, and the shaded region shows the 1 sigma region of the SPE gaussian region (\(\sigma_{\rm SPE}\)). } \label{fig:GainTemperatureResults} 
    \end{figure}

The SPE charge is parametrised by fitting a model developed in Ref.~\cite{bellamy1994_PMT}. The probability distribution function (PDF) is given in Eq.~\ref{eq:spe_q_model}, where the true SPE charge is modelled by the convolution of Gaussians with a freely floating mean, $\mu_{\rm SPE}$, and a standard deviation, $\sigma_{\rm SPE}$. Additionally, we include components to account for underamplified signals (i.e., due to missing a dynode stage during amplification) by scaling $\mu_{\rm SPE}$ and $\sigma_{\rm SPE}$ by a fraction, $\delta_{\rm uamp}$, of the fully amplified charge. 
    \begin{figure}[hb]
        \centering
        \captionsetup[subfigure]{justification=centering}
         \includegraphics[height = 6 cm]{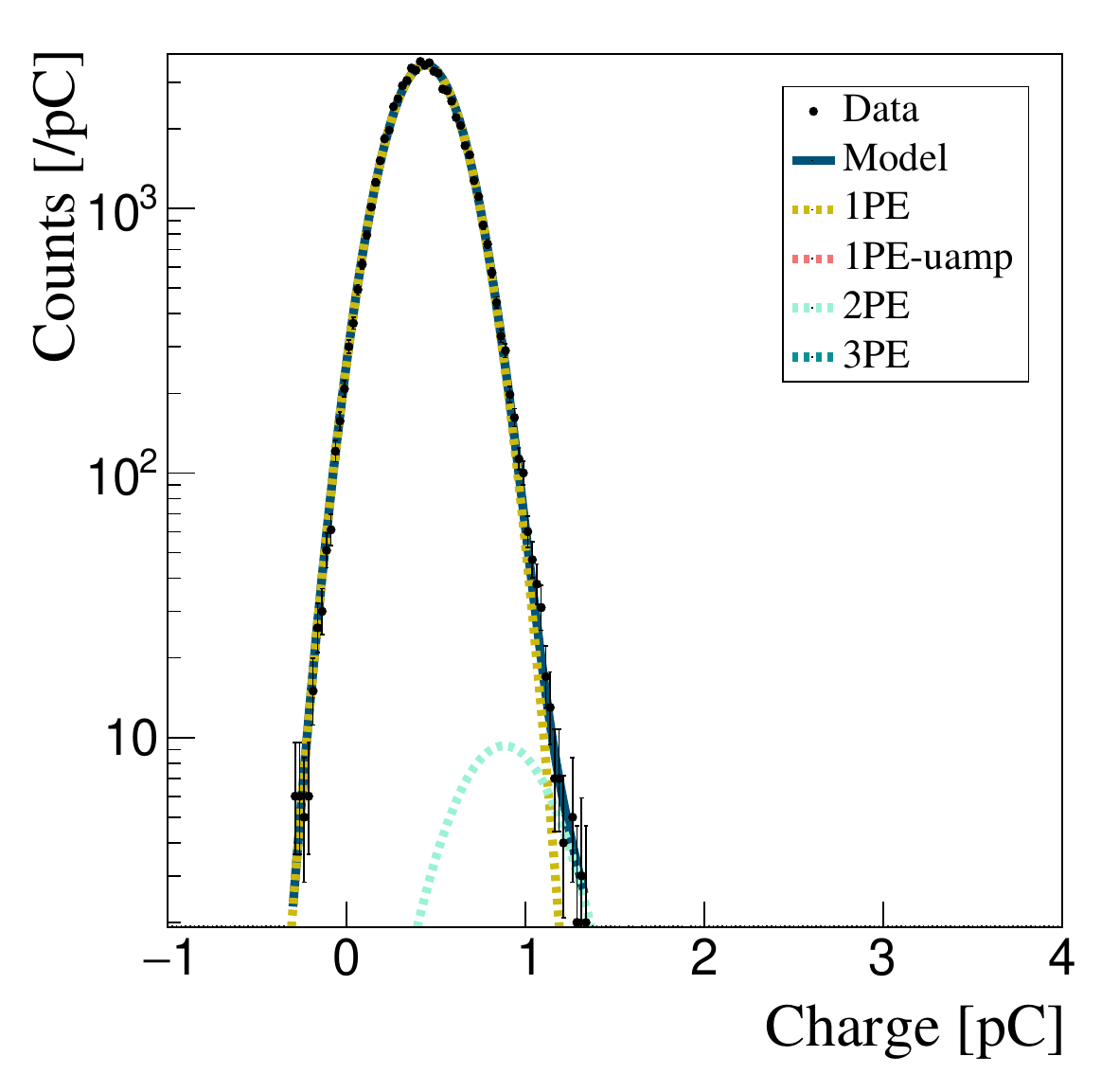}
         \includegraphics[height = 6 cm]{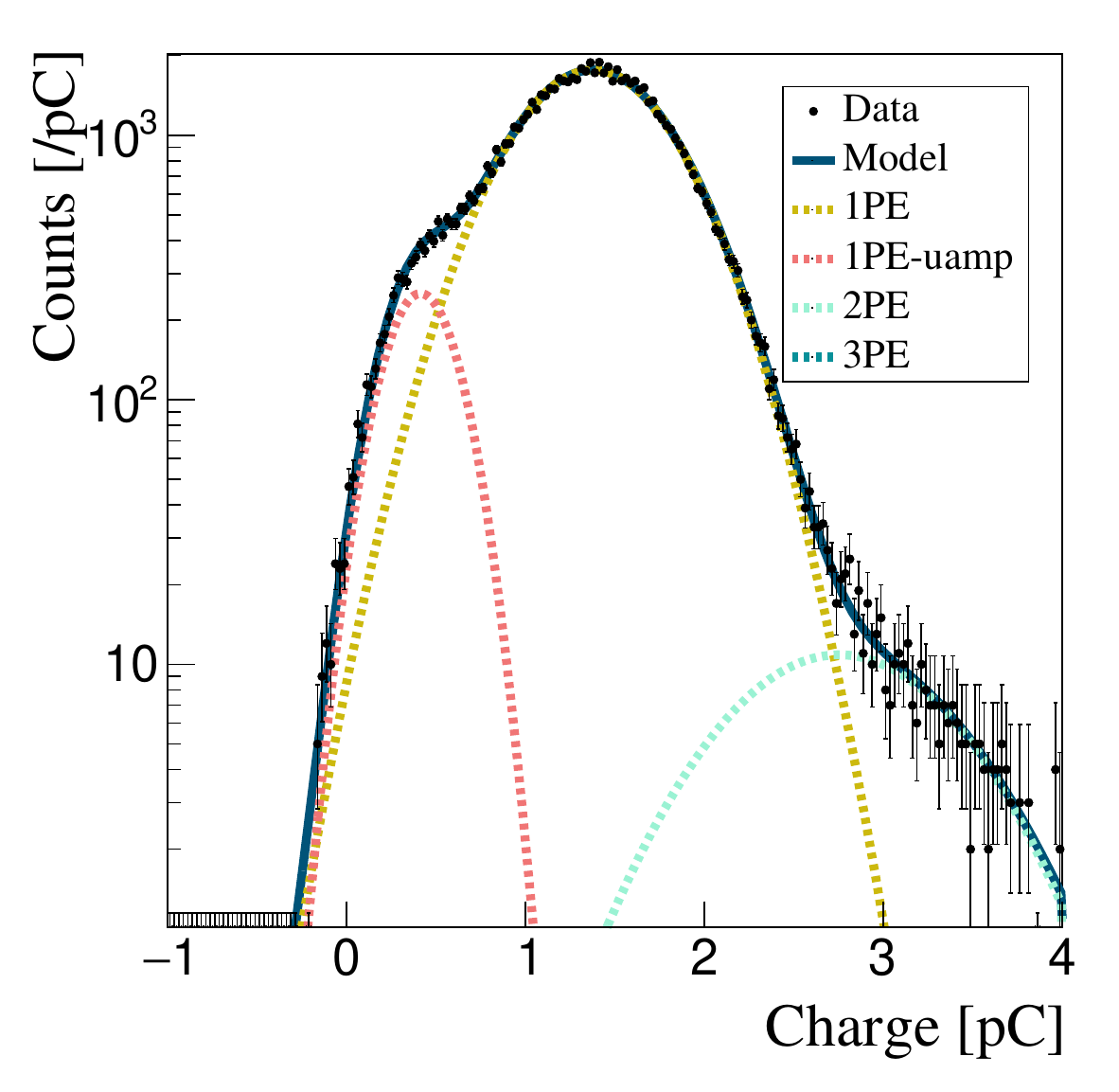}
         \label{fig:SPEFit1750V}
        
        \caption{Fitted charge spectra for PMT BC0174. At (left) 1500~V and (right) 1750~V.}
        \label{fig:BC0174_FittedLogy}
    \end{figure}
The relative proportions of the higher PE events are scaled based on the mean occupancy of the dataset~\cite{bellamy1994_PMT}. The probability that $n$-PEs are observed at a given mean $\mu_e$ is given by  
    \begin{equation}
        \label{eq:MeanOcc}
        P(n,\mu_e) = \frac{\mu_e^n e^{-\mu_e}}{n!},
    \end{equation}
where $\mu_e$ is the mean number of PEs collected at the first dynode.
We account for up to three PEs for both the fully amplified and the under-amplified components of the signal~\cite{bellamy1994_PMT}. The full PDF is of the form 
\begin{equation}
        \label{eq:spe_q_model}
        \text{PDF} = P(n,\mu_e)\otimes \left(\left(1-p_{\rm uamp}\right)\times G(\mu_{\rm SPE}, \sigma_{\rm SPE}) + p_{\rm uamp} \times G(\mu_{\rm uamp}, \sigma_{\rm uamp})\right)^{n},
\end{equation}
where $G$ are gaussian functions with mean $\mu$ and standard deviation $\sigma$. $\mu_{\rm uamp}=\mu_{\rm SPE} \times \delta_{\rm uamp}$, $\sigma_{\rm uamp}=\sigma_{\rm SPE} \times \delta_{\rm uamp}$, and $p_{\rm uamp}$ is the probability of under-amplified signals. The values of $\mu_{\rm SPE}$, $\sigma_{\rm SPE}$ ,$\delta_{\rm uamp}$, $\mu_{e}$ and $p_{\rm uamp}$ are determined based on the SPE dataset. The fit to data is shown in Fig.~\ref{fig:BC0174_FittedLogy}. The under-amplified component is scaled as a fraction of the SPE fit: in the case that the under-amplified mean is less than 0.1~pC and the SPE mean is less than 0.2~pC, this fit is repeated without the under-amplified component. The measured SPE values of these fits are shown in Table~\ref{tab:r11065_gain}. The gain ($\mu_{\rm gain}$) of the PMTs can be calculated using the relation 
\begin{equation} \label{eqn:gain}
        \mu_{\rm gain} = \frac{\mu_{\rm SPE}}{q_{\rm e}},
    \end{equation}
where $\mu_{\rm SPE}$ is the mean SPE charge per photo-electron and $q_{e}$ is the electron charge. We assume that the voltage is approximately evenly distributed for all the dynodes such that $\mu_{\rm gain}$ for a PMT with $N$ dynodes (R11065 has 12 dynodes) is given by~\cite{kaplan2014hamamatsu}
    \begin{equation}\label{eq:gainfit}
        \mu_{\rm gain} = (a \times \Delta V^{k})^{N} = a^{N} \times \left(\frac{V}{N+1}\right)^{kN} = A \times V ^{kN},
    \end{equation}
where $A$ is an arbitrary constant, $k$ is the constant for a given PMT, and $V$ is voltage in kV. The results of this fit are listed in Table~\ref{tab:r11065_gain_fit}. There is agreement between the model and our data (Fig.~\ref{fig:GainTemperatureResults}). A systematic error on the SPE values, introduced by the hardware trigger threshold is determined by comparing the results between a laser-triggered run (which has no threshold on the PMT under analysis) and PMT-triggered runs with a nominal hardware threshold of 1~mV (1.5~mV for analysis). The systematic error on the gain is determined to be $\approx 1 \%$  at all operating voltages. The mean SPE charge measured at the different voltages for each of the two R11065 (BC0174 and BC0175) PMTs tested are shown in Table~\ref{tab:r11065_gain}. This is used to convert measured charge to an estimate of the number of photoelectrons (NPE)  \( = q_{i}/\mu_{\rm SPE}\) where $q_i$ is the charge in each event.

    \begin{table}[htb]
    \centering
        \caption{Best-fit results for the R11065 PMT gain as a function of voltage. The fit model is given in Eq.~\ref{eqn:gain}  and applied to the data shown in Fig.~\ref{fig:GainTemperatureResults}.}
    \label{tab:r11065_gain_fit}
    \renewcommand{\arraystretch}{1.1} 

    \begin{tabular}{|c|c|c|}\hline
    PMT & $A$ & $k$\\
         \hline
        BC0174 & $(1.47 \pm0.12)\,\times 10^5$ & $0.604 \pm 0.013$ \\
        BC0175 & $(1.79 \pm0.11)\,\times 10^5$ & $0.624 \pm 0.010$ \\ \hline
\end{tabular}
\end{table}

\subsection{Gain calibration with dark counts}

The SABRE South NaI(Tl) detectors will be calibrated in-situ with a radioactive source system~\cite{Barberio_2025}, and through low levels of intrinsic impurities that produce measurable signals (e.g. $^{22}$Na, or $^{40}$K tagged by the veto system~\cite{COSINE-100:2024ola}). Although they provide a measure of relative efficiency and energy scale, they typically do not provide detailed information at the SPE level. The ability to monitor the PMT's gain with dark count events allows for monitoring of PMT performance over the lifetime of SABRE South, with the potential for higher frequency monitoring. Here, we demonstrate the ability for such a calibration using dark events, which are due to thermionic emission from the photocathode. 

\begin{figure}[htb]
    \captionsetup[subfigure]{justification=centering}
    \centering
        \includegraphics[height = 6cm]{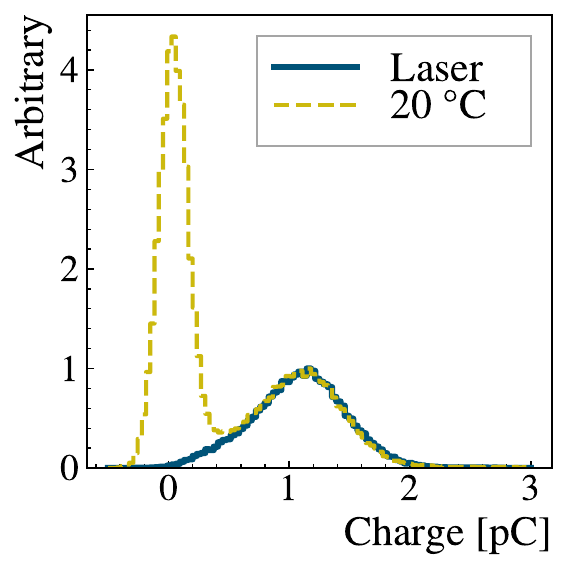}
        \includegraphics[height = 6cm]{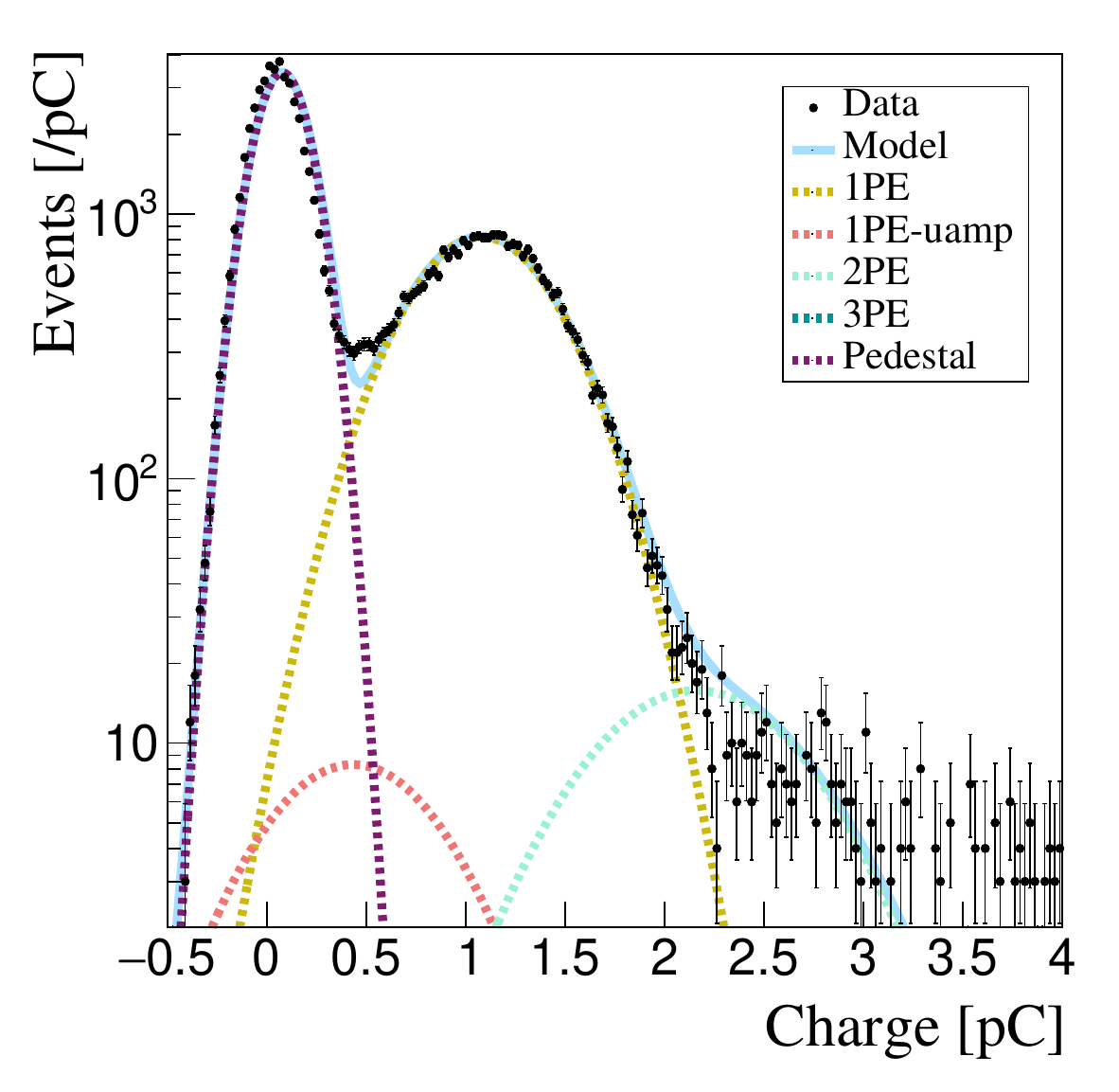}
    \caption{(left) Charge distributions of dark rate and laser triggered data used in the SPE measurement, normalised at the SPE mean value. (right) A fit to the distribution of dark rate data fitted using Eq.~\ref{eq:spe_q_model} with the addition of a Gaussian component for the pedestal.}
    \label{fig:CalibrationValidation}
\end{figure}

Validation of the gain calibration method is performed with dark count data taken using the setup in Fig.~\ref{fig:ExpSetup} (middle). Data are collected over a range of temperatures between 15$\degree$C and 30~$\degree$C (Sec.~\ref{sec:dr}). The datasets are then fitted using Eq.~\ref{eq:spe_q_model} by adding a Gaussian component for the pedestal (electrical noise events with charge of 0~pC) and requiring this component to have a mean near 0~pC (Fig.~\ref{fig:CalibrationValidation}). We find agreement between the laser and dark rate methods. At 1500~V the SPE charges for the laser dataset are measured to be $0.581\pm0.001$~pC for BC0175 and $0.440\pm0.001$~pC for BC0174. The dark rate dataset at 22~$\degree$C they are measured to be $0.585\pm0.001$~pC for BC0175 and $0.432\pm0.002$~pC for BC0174 (shown in Fig.~\ref{fig:GainTemperatureResults}). 

\subsection{Timing} \label{sec:timing}
\begin{figure}[htb]
    \centering
    \includegraphics[height=6cm]{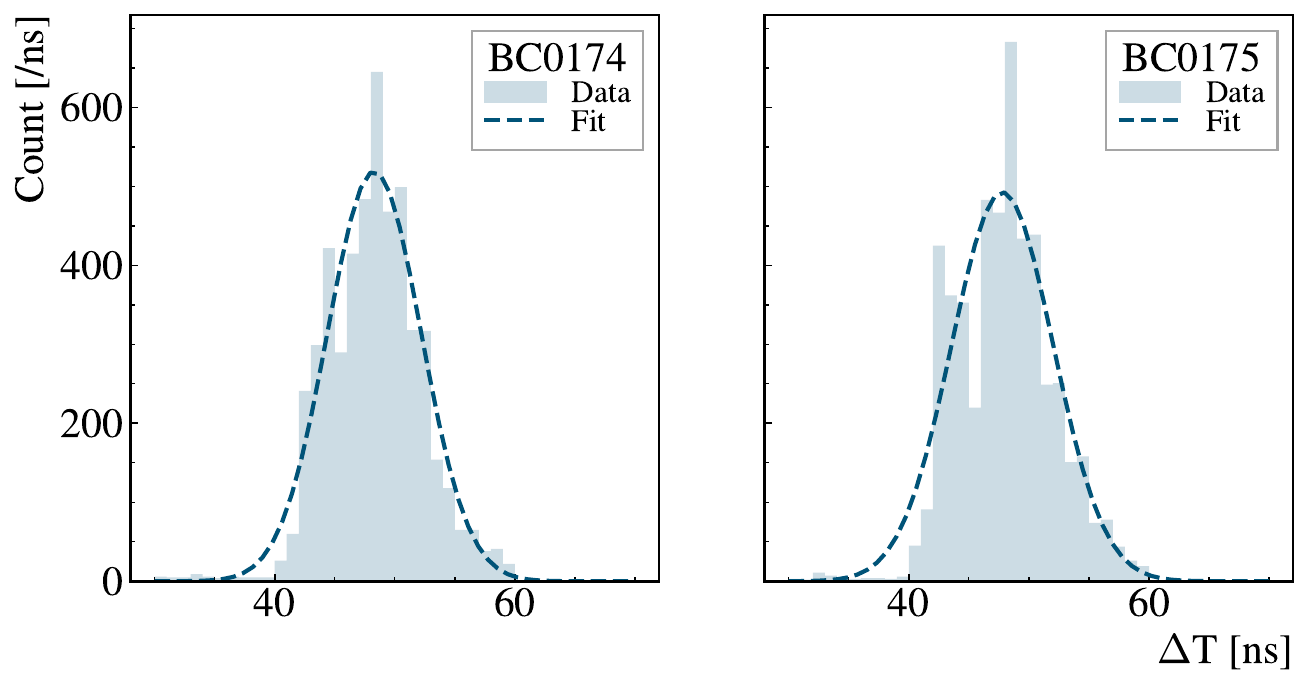}
    \caption{Difference in leading edge times (Equation~\ref{eq:tdiff}) for PMTs BC0174 (left) and BC0175 (right). The time-difference distributions are fit with a single Gaussian model.} 
    \label{fig:transittime-dists}
\end{figure}

    The measurement of the response time of each PMT is used to correctly model signals in detector simulation, and as a quality control measure. The response time of the PMTs is characterised by the transit-time distribution (TTD), which is typically modelled as a Gaussian function, the mean being the transit time (TT) and the width corresponding to the transit-time spread (TTS). The TT is the time between the photon striking the photocathode and the electron cascade detection on the anode. This is measured using an HPK S10362-11-050 silicon photomultiplier (SiPM) as a reference detector, which has a TT of 100 to 500~ps~\cite{PUILL2012354} and a TTS of 200 to 300 ps as specified by HPK. The SiPM signal is used as a coincidence trigger source. The single photon pulse properties are measured using the SPE data-taking with the PMT perpendicular to the laser swapped for an SiPM. The R11065 PMT is operated at 1500~V and the acquisition conditions are otherwise identical to the measurement in Sec.~\ref{sec:spe}. The data presented are required to have a charge within 0.1~pC around the mean SPE value (chosen conservatively to ensure a high purity SPE sample) and coincident signals from the laser, PMT, and SiPM. The start times of the pulses in the SiPM and PMT are identified using a leading-edge trigger with a threshold of 1.5~mV. The difference in leading edge times (between the SiPM and PMT) is used to characterise the mean and spread of the TTD,
    \begin{equation} \label{eq:tdiff}
        \Delta T = T_{\rm PMT} - T_{\rm reference},
    \end{equation}
    where the reference here is the SiPM. The distribution of these timing differences is shown in Fig.~\ref{fig:transittime-dists}. The PMT-SiPM time difference is fitted with a Gaussian. The results are reported in Table~\ref{tab:transit_time}, where the nominal values quoted by HPK are TT, 46~ns and the TTS, 9~ns.
    
    \begin{table}[htb]
        \centering
        \caption{The transit-time mean and spread of the R11065 measured using the SiPM as the reference sensor. The displayed error comes from the fits, and systematics arises from SiPM timing and resolution.}
        \label{tab:transit_time}
        \renewcommand{\arraystretch}{1.1} 
        
        \begin{tabular}{|c|c  |c |}
        \hline
 & \multicolumn{2}{c|}{\textbf{T$_{\rm PMT}$ - T$_{\rm SiPM}$}} \\
        \hline
             \textbf{PMT} & \textbf{TT} [ns] & \textbf{TTS} [ns]  \\
             \hline
             \textbf{BC0174} & 48.3 $\pm$ 0.1$_{\rm fit}$ $\pm$ 0.5$_{\rm syst.}$& 9.2 $\pm$ 0.1$_{\rm fit}$ $\pm$ 0.3$_{\rm syst.}$\\
             \textbf{BC0175} & 47.8 $\pm$ 0.1$_{\rm fit}$ $\pm$ 0.5$_{\rm syst.}$& 10.0 $\pm$ 0.1$_{\rm fit}$ $\pm$ 0.3$_{\rm syst.}$\\
             \hline
        \end{tabular}
    \end{table}
  
\section{PMT Background}

In this section, we present measurements of the characteristic backgrounds within the PMT. This includes measurements of the dark rate and the after-pulse probability. These background sources can potentially result in a fluctuating background component in the low energy range where dark matter events are expected.

\subsection{Dark Rate} \label{sec:dr}
    The dark rate of the R11065 PMTs must be measured as a function of both the operating voltage and the temperature. This is key to developing an accurate model of the background caused by the PMTs near the DAMA/LIBRA RoI. If this effect is not correctly modelled, then fluctuations in the PMT temperature and gain can potentially cause biases in the annual modulation search. The expected effect of temperature on the dark rate is primarily due to thermionic emission from the photocathode. Adapting the model used by HPK, we model the dark rate as 
    \begin{equation}\label{eq:dr_model}
        R = A \cdot T^{5/4}\cdot e^{-q_e \psi/KT} + C,
    \end{equation}
    where $T$ is the absolute temperature of the cathode, $q_e$ is the electron charge, $\psi$ is the cathode work function, $K$ is the Boltzmann constant and C is a constant to account for  electronic background~\cite{kaplan2014hamamatsu}.

    This background is significant for SABRE South, as the expected operating temperature of SUPL is \supltemp{}, while the natural temperature of the surrounding rock is approximately $\sim35~\degree$C. SUPL will be cooled constantly, as the surrounding rock is approximately $\sim35~\degree$C, fluctuations in temperature may be introduced by changes in cooling performance and other operations in SUPL. Short-term fluctuations in laboratory temperature are unlikely to have an effect, as the liquid scintillator and passive shielding provide tens of tonnes of thermal bulk between the PMTs and the air in SUPL, however long term temperature changes could induce a modulation.

    \begin{figure}[htb]
            \captionsetup[subfigure]{justification=centering}
            \centering    
            \includegraphics[height = 6 cm]{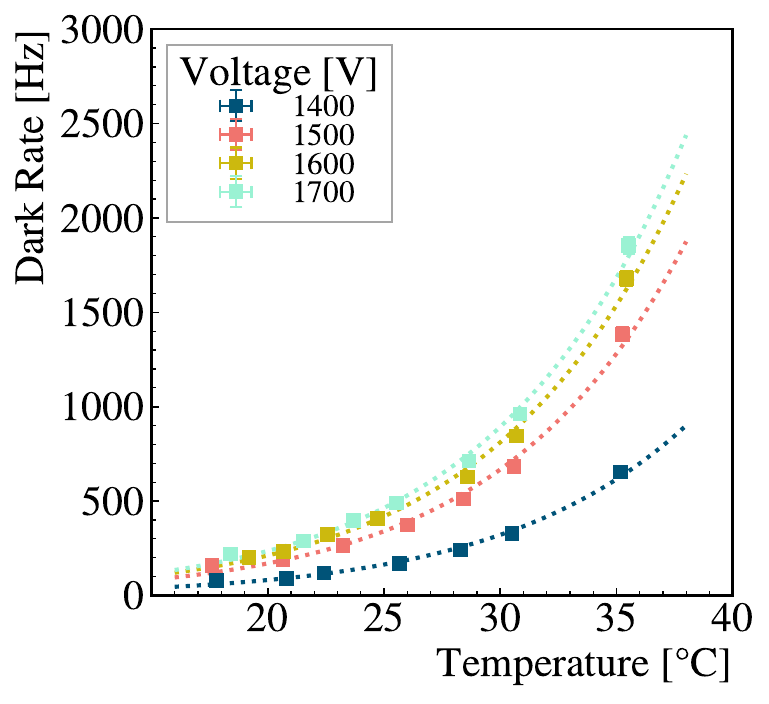}
            \includegraphics[height = 6 cm]{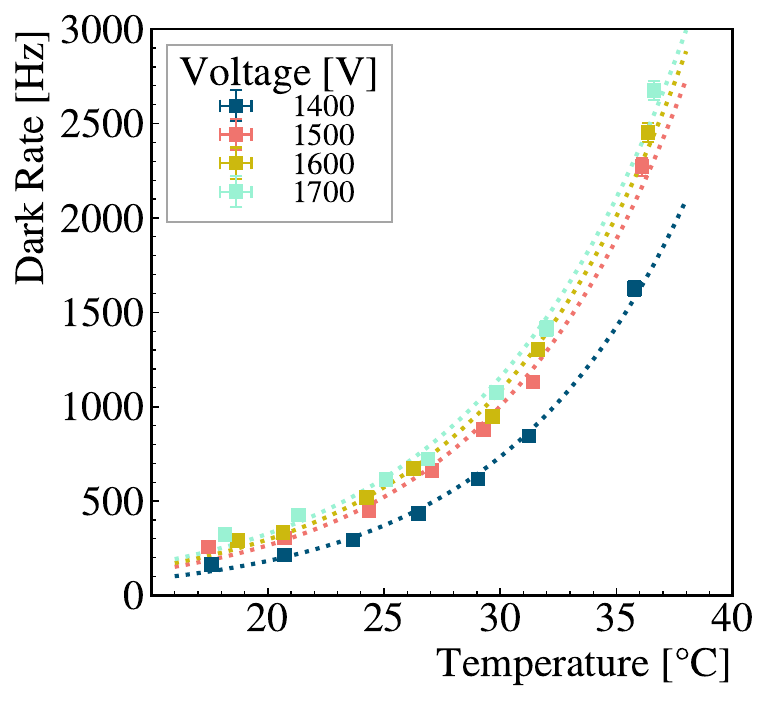}
            \caption{Dark Rate as a function of temperature for BC0174 (left) and BC0175 (right) fitted with the HPK model Eq.~\ref{eq:dr_model}. Dark Rate events must pass a threshold of 1.8 mV and a charge threshold of 0.5~$\times$~SPE charge. The results are reported at four voltages: 1400~V, 1500~V, 1600~V, and 1700~V from lowest to highest measured dark rates. The errors on the temperature are due to systematic errors on the RTD sensors.}
            \label{fig:DR_func_temp}
        \end{figure}

The temperature in the PMT thermal testing chamber is controlled by a digital thermostat that powers the heating and cooling systems. Additional temperature monitoring is provided by a prototype of the SABRE South slow control system detailed in Ref.~\cite{Krishnan:2021yqc} using RTD sensors. The thermal testing system is capable of maintaining the internal temperature to within 0.1$\degree$C of the set temperature for temperatures above $15~\degree$C. The PMTs are placed in the dark box and powered up to the lowest voltage of the run at least 12~hours prior to the initial measurement to ensure the voltage has stabilised. At each voltage step an additional 30~minutes was given to ensure stable operation. The measurement is carried out with a threshold of 1.2~mV which is just above the fluctuation of the baseline. The measurement was taken to collect at least 100 000 dark events in each combination of voltage and temperature. The measurements were taken at 100~V increments in a temperature range from 17.5~--~35~$\degree$C  in 2.5~$\degree$C increments, to provide detailed information well above and well below the planned operating temperature of SUPL. The results of these measurements are shown in Fig.~\ref{fig:DR_func_temp} and fit with the model given in Eq.~\ref{eq:dr_model} floating the normalisation, work function and background value. Noting that electrical noise contribution is independent of temperature. The results show good agreement with the adapted model for the dark rate. The error on the temperature is calculated based on Eq.~\ref{eq:RTD_T_Err}. The fitted work function ranges from 1.17~--~1.3 eV, this is lower than those measured by other studies on HPK PMTs using a bialkali cathode typically ranging from 1.2~--~1.5~eV \cite{Wilson:2023wdf}. Further work is needed to understand this discrepancy.

\subsection{Light Emission rate}

\begin{figure}
    \centering
    \includegraphics[height=6 cm]{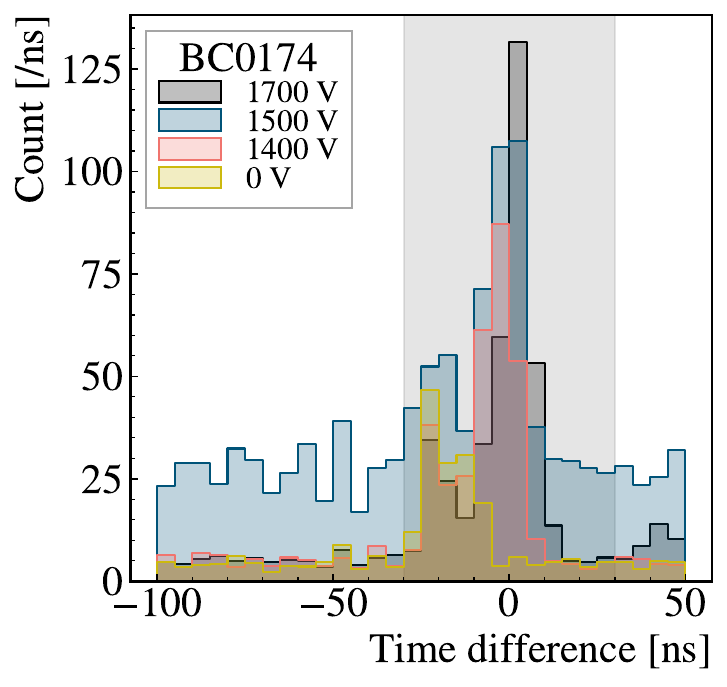}
    \includegraphics[height=6 cm]{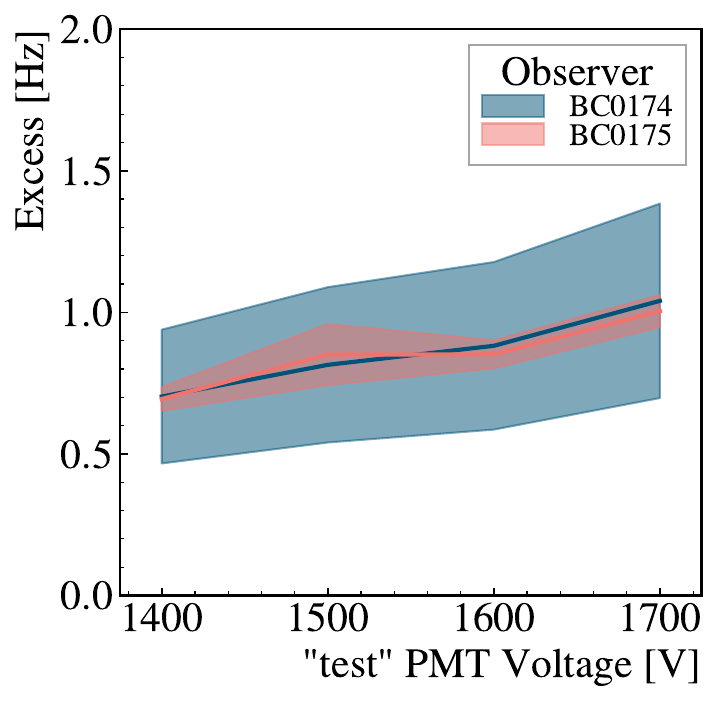}
    \caption{(Left) The time difference between the observer PMT and the "test" PMT, the different colours are the operating voltage of the measured PMT, BC0174. (Right) The observer PMT excess trigger rate, at different operating voltage of the observer. The band shows the statistical error accounting for the elevated background in the unbiased case.}
    \label{fig:CorrelatedLightEmission}
\end{figure}

Spontaneous emission of light from the internal PMT structure has been identified and measured by studies of the HPK R11410 PMTs~\cite{Antochi_2021}. The exact cause of such events is uncertain, but these present large background contributions and can have a significant impact on the overall PMT background of the detector. For SABRE South where the PMTs face one another via a NaI(Tl) crystal, the emission of this light can potentially cause a significant background. A measurement of this background is performed using the SABRE South mock setup shown in Fig.~\ref{fig:ExpSetup} (right), where the PMT windows are separated by a gap of 80 mm to avoid damage. The setup has one observer PMT operating at 1500~V and a secondary (test) PMT which has its voltage varied from 1400~V~--~1700~V. A baseline measurement of the dark rate of the observer is taken with the test PMT at 0~V. The excess rate is measured after applying a series of selection criteria: peak height~$>~2.8$~mV, charge~$>~0.5$~PE, and the time difference between pulses in the two PMT must be within 60~ns of one another. The elevated rate in the baseline case is caused by electrical noise picked up by the signal cables. The peak observed at around -25~ns is a result of cross-talk between the PMT cables, where a large pulse in one PMT can induce a much smaller, but non-negligible signal in the nearby signal cable. A sideband subtraction technique is used to remove random background, with a width of 800~ns (-1000~--~-200~ns), which is scaled to the same size as the coincidence window. After the sideband subtraction we account for the elevated background rate in the unbiased case, by subtracting the baseline case from the biased PMT data sets. The time difference distribution and the excess rate results are shown in Fig.~\ref{fig:CorrelatedLightEmission}. An increase in rate of $\cal{O}$(1~Hz) rate is observed in both PMTs.The errors from the subtraction of the sideband and the unbiased case are included in the error bands.


\subsection{After-pulse Rate}

After-pulsing occurs when an electron emitted from the photocathode ionises residual gas within the PMT vacuum. Consequently, the ion travels back to the photocathode and photoelectrons are emitted that follow the amplification chain. These are observed at a later time within the waveform. Accidental coincidences of these events that originate from any pair of PMTs can imitate the DM signal. Furthermore, these after-pulses are potentially time dependent as the concentration of ions within the residual gas could fluctuate throughout the operation of the experiment. Tracking these fluctuations also allows us to understand if there are defects within our PMTs. As the crystal enclosures will be flushed with nitrogen, we may observe an increase in the after-pulse rate due to N$_2^+$ ions. The time difference between the main pulse and the after-pulses is dependent on ion species and can be calculated using the expression~\cite{Barrow_2017} 
\begin{equation}\label{eq:afterpulse_timediff}
    \Delta t = 1.134 \sqrt{\frac{L^2}{V_0}\frac{M}{Q}}.
\end{equation}
Here $L$  is the distance between the focusing grid and the photocathode in cm (4.1 cm for the HPK R11065), $V_0$ is the voltage in volts of the first dynode and $M/Q$ is the dimensionless mass-charge ratio of the ion. A list of these ions and their associated times is given in Table~\ref{tab:IonisationTimes} for the HPK R11065 PMT. 
The structural design of the HPK R11065 is the same as that of the HPK R11410~\cite{HamSpecSheet} studied in Ref.~\cite{Barrow_2017}.
    \begin{table}[htb]
        \centering
        \caption{Time differences between after-pulses and trigger pulses, for different ions in the R11065 PMT~\cite{Barrow_2017}.} \label{tab:IonisationTimes}
            \renewcommand{\arraystretch}{1.1} 
        \begin{tabular}{|l|cccc|}
        \hline
         & \multicolumn{4}{l|}{Expected $\Delta T$ for various ions [$\mu$s]}    \\ 
        \hline
        Voltage [V]& \multicolumn{1}{l|}{H$^{+}$ (M/Q = 1)} & \multicolumn{1}{l|}{He$^{+}$ (M/Q = 4)} & \multicolumn{1}{l|}{CH$_4$$^{+}$ (M/Q = 16)} & N$_2$$^{+}$ (M/Q = 28) \\ \hline
        1500            & \multicolumn{1}{l|}{0.28}         & \multicolumn{1}{l|}{0.52}          & \multicolumn{1}{l|}{1.03}            & 1.37           \\ \hline
        1600            & \multicolumn{1}{l|}{0.25}         & \multicolumn{1}{l|}{0.50}          & \multicolumn{1}{l|}{1.00}            & 1.32           \\ \hline
        1700            & \multicolumn{1}{l|}{0.24}         & \multicolumn{1}{l|}{0.48}          & \multicolumn{1}{l|}{0.97}            & 1.28           \\ \hline
        \end{tabular}
    \end{table}

\begin{figure}[htb]
    \centering
    \includegraphics[height=6 cm]{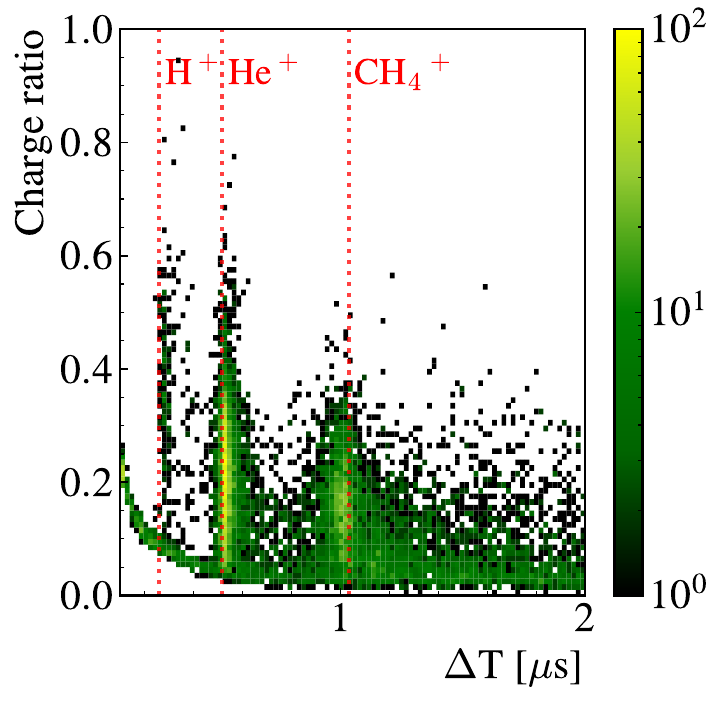}
    \caption{Charge ratio of the initial pulse with after-pulse against the time difference between the main and after-pulse peak. The dotted red lines identify the predicted time differences with respect to the primary peak.}
    \label{fig:AFP_charge}
\end{figure}

Making use of the SPE setup but running at higher laser power, we can characterise the production of after-pulse events. The main peaks of the collected events have a charge between 30 and 80 PE (integrated from -20~--~80~ns with respect to the pulse start) shown in Fig.~\ref{fig:AFP_charge}. Peaks within the waveforms are identified using ROOT TSpectrum~\cite{MORHAC2000108}, where, to identify a peak, the average charge should be above a 1.5 mV threshold in a time window of at least 3~$\times$ the signal duration (3~ns). The pulse start times are determined with a leading edge threshold algorithm, and the time difference between pulses is evaluated as in Eq.~\ref{eq:tdiff}. 
The ratio of charges in the after-pulse and in the main pulse is shown versus the time difference in Fig.~\ref{fig:AFP_charge} where the dotted lines represent the expected after-pulse times.  The after-pulse rate is determined by the number of observed events with an after-pulse to main pulse charge ratio $\geq$ 0.15 and within 100 ns of the expected after-pulse time ($\hat t$). The observed after-pulse rate is then calculated with the following expression  
\begin{equation*}\label{eq:afterpulsePerc}
    \text{After-pulse rate} = \frac{N_{(\hat{t} - 100\text{~ns}~,~ \hat{t}+100\text{~ns})}}{N_{total}\times \hat{q}_{\rm total}}
\end{equation*}
where $N_{(\hat{t} - 100~,~ \hat{t}+100)}$ is the number of events within a time window of 100 ns of the expected after-pulse time. The total number  of events ($N_{total}$) is the number of events in the PMT coincident with a laser signal  and the mean charge of the distribution $\hat{q}_{total}$. The results of this analysis are summarised in Table~\ref{tab:Aft_percentage}. The results showed that there is an increase in after-pulsing rate with voltage, this has been observed by other studies,~\cite{AKCHURIN2007121} and attributed to an increase in the kinetic energy of the ions. 
 
\begin{table}[htb]
    \centering
    \caption{The proportion of observed after-pulses following a laser induced event (Equation~\ref{eq:afterpulsePerc}). The errors displayed are the statistical error of the datasets. }
     \label{tab:Aft_percentage}
    \renewcommand{\arraystretch}{1.1} 
    \begin{tabular}{|l|c|ccc|} \hline
     &                    & \multicolumn{1}{c}{1500 V} & \multicolumn{1}{c}{1600 V} & 1700 V    \\ \hline
     &H$^{+}$ [/PE] $\times 10^{-4}$& 0.48$\pm$0.03& 0.58$\pm$0.04& 1.11$\pm$0.10\\
     BC0174&He$^{+}$ [/PE] $\times 10^{-4}$& 5.57$\pm$0.11& 6.00$\pm$0.11& 7.25$\pm$0.26\\
     &CH$_{4}^{+}$ [/PE] $\times 10^{-4}$& 2.55$\pm$0.07& 2.92$\pm$0.08& 3.92$\pm$0.02\\ \hline
 & H$^{+}$ [/PE] $\times 10^{-4}$& 0.42$\pm$0.03& 0.53$\pm$0.03& 1.26$\pm$0.04\\
 BC0175& He$^{+}$ [/PE] $\times 10^{-4}$& 5.26$\pm$0.10& 6.25$\pm$0.11& 6.88$\pm$0.11\\
 & CH$_{4}^{+}$ [/PE] $\times 10^{-4}$& 2.59$\pm$0.07& 3.04$\pm$0.07& 3.62$\pm$0.08\\
 \hline
    \end{tabular}
\end{table}

\section{Noise~--~Scintillation Discrimination via Machine Learning}
The detector thresholds of NaI(Tl) experiments are limited by the LY of the crystal module and the coincident background rate of the PMTs. Efforts on reducing the coincident background have been focused on improved analysis techniques using machine learning. In this section, we employ a boosted decision tree (BDT) model to capture complex non-linear relationships between variables,for the discrimination between signal and background events. BDTs offer a balance between classification performance and interpretability, making them a practical alternative to more opaque models such as neural networks. Here we use variables that have previously been used in studies by SABRE~\cite{SABRE_NAI33Characterisation}, ANAIS-112~\cite{Coarasa_2022}, COSINE-100~\cite{COSINE-100:2020wrv}, and DAMA/LIBRA~\cite{DAMA/LIBRA_apparatus}. The study is performed using data collected in a controlled test environment, with and without a crystal. The performance of this model is assesed by applying this model on alternate PMTs, as well as making a comparison to a simpler approach using only the parameters developed by DAMA/LIBRA.

\subsection{Data}

Using the coincidence setup described in Sec.~\ref{sec:exp_setup} tagged signal and background data is collected with the two PMTs (BC0174 and BC0175), where the gains are matched. For the background data, the crystal is not present and an air gap of 80 mm (the length of the encapsulated EPICS crystal) is maintained between the two PMTs. The signal data set is taken with the crystal and a $\gamma$ (\Ba{}) source to provide scintillation signals. These data sets are obtained by requiring a coincidence between at least two of the four detectors (the two R11065 PMTs and two plastic scintillator panels) that pass hardware thresholds of 1.5~mV for the R11065 PMTs and 10~mV for the plastic scintillator PMTs. The trigger requires a coincidence window of 240 ns. This configuration has a LY of $\approx$ 4~PE/keV (combined between the two PMTs), which is 30\% of the typical LY of a SABRE crystal detector. The LY is relatively low because of the lack of coupling material between the PMT and the crystal, leading to internal reflection. This analysis is done in units of the number of photoelectrons (PE) to be agnostic to the LY, which will vary for the final SABRE crystal detector modules~\cite{SABRE_NAI33Characterisation}.  The two PMTs are gain-matched for this study. This study uses a \Ba{} source with 80~keV $\gamma$ ray produced via electron capture. This source provides a low energy scintillation source appropriate for studying the 1~--~6~\kevee{} (10~--~60 PE) RoI. 

\subsection*{Coincidence Filter}
\begin{figure}[htb]
    \captionsetup[subfigure]{justification=centering}
        \centering
        \includegraphics[height=5 cm]{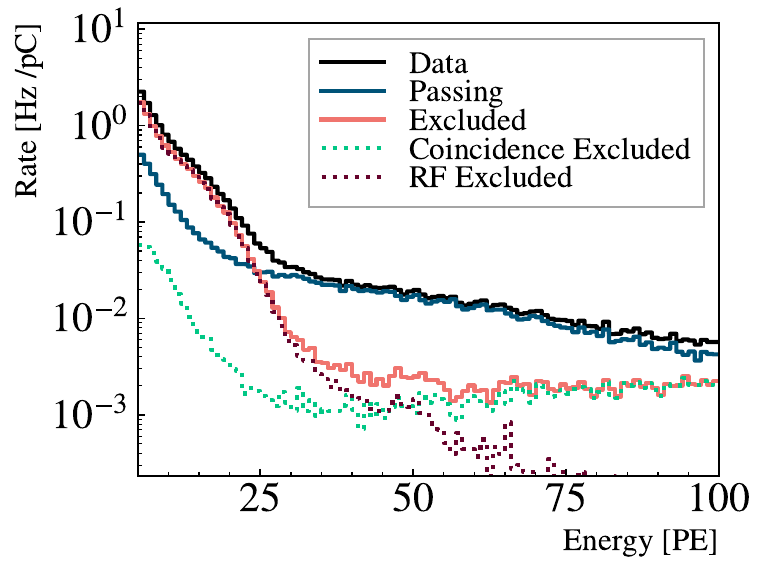}
        \includegraphics[height=5 cm]{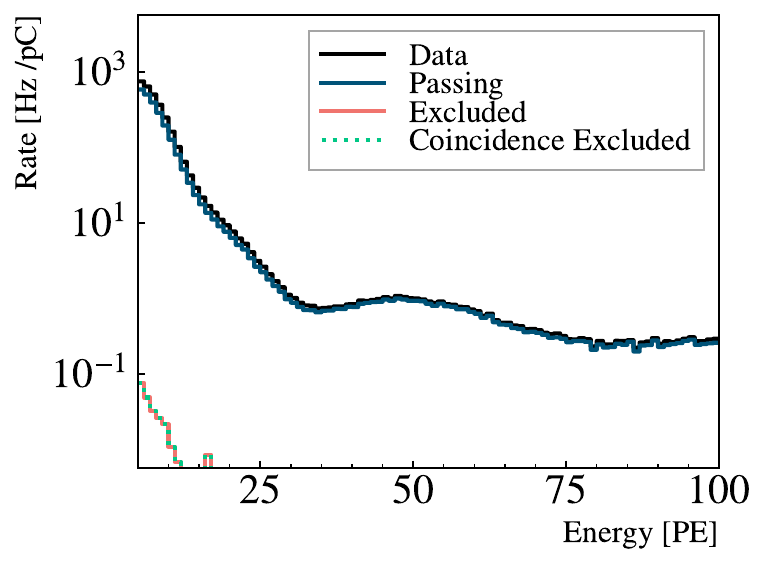}
    \caption{Number of photo-electron (NPE) distributions for the background dataset (left) and the \Ba{} signal dataset (right). Displays the full collected data, the events excluded from the coincidence and muon criterias as well as the events excluded from the RF criterias. The remaining passing events are also displayed.}
    \label{fig:energyspectrum}
\end{figure}

The initial data set has a population of 2 (6.1) million background (signal) events taken over 720 (33)~minutes. To reduce PMT-induced noise, a coincidence signal between the two PMTs is required. Hence, we require candidate events to be observed in both PMTs. 
Here we require that every event in each PMT passes a software peak height threshold of 1.5~mV ($\approx$0.3 SPE peak height) chosen based on the SPE analysis, and we choose a coincidence window of 200 ns to account for the long scintillation time of the NaI(Tl) crystal. We additionally make use of the plastic scintillators to tag and veto the cosmogenic muon events. These muon events are excluded and categorized along with other coincidence-related cuts as "coincidence excluded" in Fig.~\ref{fig:energyspectrum}, which rises above 75 PE. The impact of these cuts on the signal dataset is minimal due to the significantly shorter data-taking period.

\subsection*{RF Filter}

RF interference can readily affect the signal quality of our system and mask some of the obtained data sets. This can be caused by systems on the same electrical circuit, and can be difficult to isolate. When caused by external systems we expect to observe a random increase then decrease in our data taking rate, corresponding to devices being turned on and off. In our data we found variations within the trigger rate of our system, suggesting the presence of RF interference. Introduction of a pre-amplifier would allow us to separate electronic noise from real scintillation events, but would lead to a limitation in the dynamic range of the PMT. Therefore a pre-amplifier was not used during these tests but is currently being tested for the final experiment~\cite{SABRE_tdr}. The observed noise for our data is most likely caused from the building's electrical circuit. Additional measures are being taken for SUPL to limit this effect. 

\begin{figure}[htb]
    \centering
    \includegraphics[width = 0.95\columnwidth]{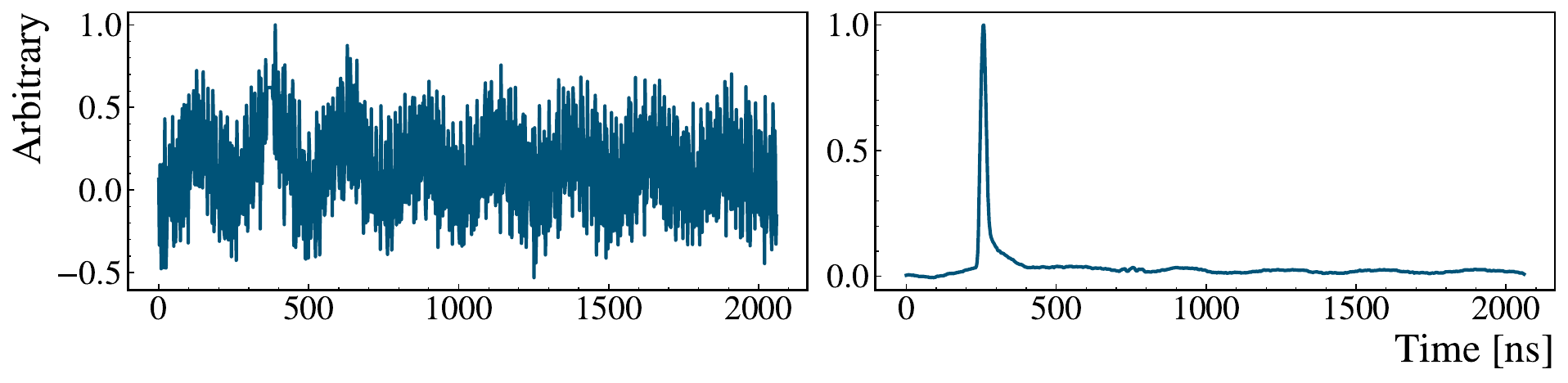}
    \caption{Noise and Signal templates, where the x-axis is the time within an event, and the amplitudes have been normalised to 1.}
    \label{fig:TemplateRF} 
\end{figure}

For this analysis we employ a software approach. To to do this we calculate a score ($\mathcal{L}_{\rm{RF}}$) by comparing each event with noise and PMT templates (Fig.~ \ref{fig:TemplateRF}) following a log-likelihood calculation,
\begin{equation}
\label{eq:Loglikelihood}
    {\rm{ln}} \mathcal{L} = \sum_{i} \left[T_{i} - W_{i}\left(1+{\rm{ln}}\left(\frac{W_{i}}{T_{i}}\right)\right)\right].
\end{equation}
Here $T_{i}$ and $W_{i}$ are the $i^{\rm th}$ entries for the template and waveform, respectively. Each waveform is then normalised and compared with the template, and a relative similarity score is calculated based on the difference between the likelihood score for each event and the two templates,\begin{equation}
    \label{eq:pl}
    \mathcal{L}_{\rm{RF}} = \frac{\rm{ln}~\mathcal{L}_{\rm{noise}} - \rm{ln}~\mathcal{L}_{\rm{PMT}} }{\rm{ln}~\mathcal{L}_{\rm{noise}} + \rm{ln}~\mathcal{L}_{\rm{PMT}} }.
\end{equation}
The noise template is created from data taken using a non-terminated cable, this data collects only electronic background and won't have any PMT background. This is averaged to form the template for RF backgrounds. To form the PMT template we select events from our dark dataset requiring a charge $\geq 1$~pC ($\approx 2 $ SPE) which is again averaged to form the signal template. 

\begin{figure}[htb]
    \centering

    \includegraphics[height=6cm]{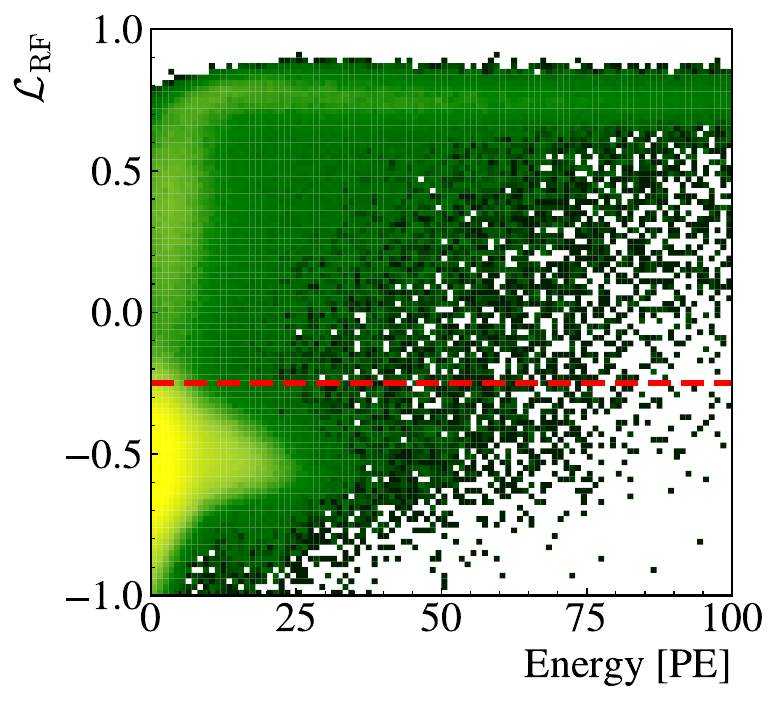}
    \includegraphics[height = 6cm]{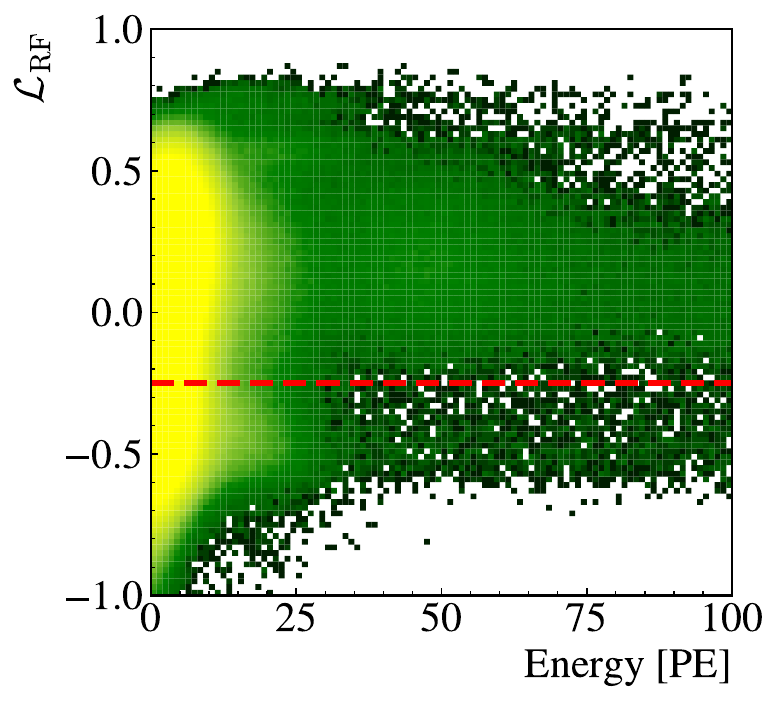}
    \caption{$\mathcal{L}_{\rm RF}$ versus approximate number of photoelectrons for test samples of RF noise (left) and signal events (right). An analysis threshold is placed at $\mathcal{L}_{\rm{RF}}$$\geq$ -0.25 (dashed red line). }
    \label{fig:pl_v_e}
\end{figure}

The likelihood scores versus charge distributions for both the signal and noise datasets are shown in Fig.~\ref{fig:pl_v_e}. There are two separate populations in the background dataset: the RF noise centred around $\mathcal{L}_{\rm{RF}}=$~-0.5 and thermal emission noise near $\mathcal{L}_{\rm{RF}}=0.8$. In the signal dataset most events are centred around $\mathcal{L}_{\rm{RF}}=0$ at higher charge; this is because the signal template is based on SPE-like events (a single large peak). In the subsequent study, events with $\mathcal{L}_{\rm{RF}}$$\leq$~-0.25 are removed prior to training and the determination of classifier performance.

\subsection*{Variables}

The variables used for the BDT model are based on those designed to be applied to a combined signal from pairs of PMTs. These variables were developed to distinguish between PMT (electronics and thermal) noise and scintillation events. These methods primarily rely on the scintillation characteristics of NaI(Tl), which exhibits a fast rise time of 50~ns followed by a long decay time of 250~ns. Taking the ratio of the charge in the fast and slow components to the total event charge provides good separation for higher-energy events ($>$2 keV\textsubscript{ee}), as demonstrated by DAMA/LIBRA~\cite{DAMA/LIBRA_apparatus}. However, this approach has proven insufficient for achieving sensitivity below 1~keV\textsubscript{ee}, where PMT electronic noise remains one of the primary limitations in \nai{}-based experiments.

The DAMA/LIBRA experiment separates noise and scintillation events by taking advantage of these fast and slow components that define the variables X1 and X2~\cite{DAMA/LIBRA_apparatus}. These variables seek to identify the main charge deposition region within the signal by looking at the first 50~ns of an event and the last 500~ns, defined as:
\begin{equation}\label{eqn:DAMAX1}
    \text{X1} = \frac{\sum_{t=100 \text{ ns}}^{t=600 \text{ ns}}\rm{S}_{0}(t) + \rm{S}_{1}(t)}{\sum_{t=0 \text{ ns}}^{t=600 \text{ ns}}\rm{S}_{0}(t) + \rm{S}_{1}(t)},
\end{equation}
and
\begin{equation}\label{eqn:DAMAX2}
    \text{X2} = \frac{\sum_{t=0 \text{ ns}}^{t=50 \text{ ns}}\rm{S}_{0}(t) + \rm{S}_{1}(t)}{\sum_{t=0 \text{ ns}}^{t=600 \text{ ns}} \rm{S}_{0}(t) + \rm{S}_{1}(t)}.
\end{equation}

Where $\text{S}_0$ and $\text{S}_1$ are the signals from the 0$^{\rm th}$ and 1$^{\rm st}$ PMTs respectively. The ANAIS-112 collaboration~\cite{Coarasa_2022}, has generalised these formulae into the charge-accumulated pulse (CAP) variables, defined as
\begin{equation}\label{eqn:CAPx}
    \text{CAP}_x = \frac{\sum_{t=t_{0} \text{ ns}}^{t=x \text{ ns}}\rm{S}_{0}(t) + \rm{S}_{1}(t)}{\sum_{t=t_{0} \text{ ns}}^{t=t_{max} \text{ ns}}\rm{S}_{0}(t) + \rm{S}_{1}(t)}.
\end{equation}
These variables can be defined over smaller subregions to investigate lower PE events. For this work, we use t$_{\rm max}$ = 1.2 $\mu$s and t$_0$ = -0.1 $\mu$s from the pulse start. Another variable commonly used in the field is the charge-weighted mean time, 
\begin{equation}\label{eqn:MeanTime}
    \log\left(\mu_p\right) = \log\left(\frac{\sum_i \rm{A}_i \rm{t}_i}{\sum_i \rm{A}_i}\right),
\end{equation}
where A$_{\rm i}$ is the amplitude and t$_{\rm i}$ is the time in the i$^{th}$ bin of the waveform. Additional parameters studied in the training are the skewness and kurtosis, which describes the degree of symmetry of the waveform and the size of the tail in a waveform and the $\mathcal{L}_{\rm{RF}}$ parameter, used to remove RF noise. Lastly, a peak finder is implemented with the ROOT TSpectrum algorithm~\cite{MORHAC2000108}. This algorithm estimates the background within a waveform to identify the peaks via a deconvolution process. It is configured to run with a peak height threshold of 1.5 mV and a minimum pulse width of 3 ns, chosen based on the width of an SPE signal and provides the number of peaks within an event (denoted "N$_{p}$").

\subsection{Selection}\label{sec:PMT_bdt_selection}

\begin{figure}[htb]
    \captionsetup[subfigure]{justification=centering}
         \centering
         \includegraphics[height = 6cm]{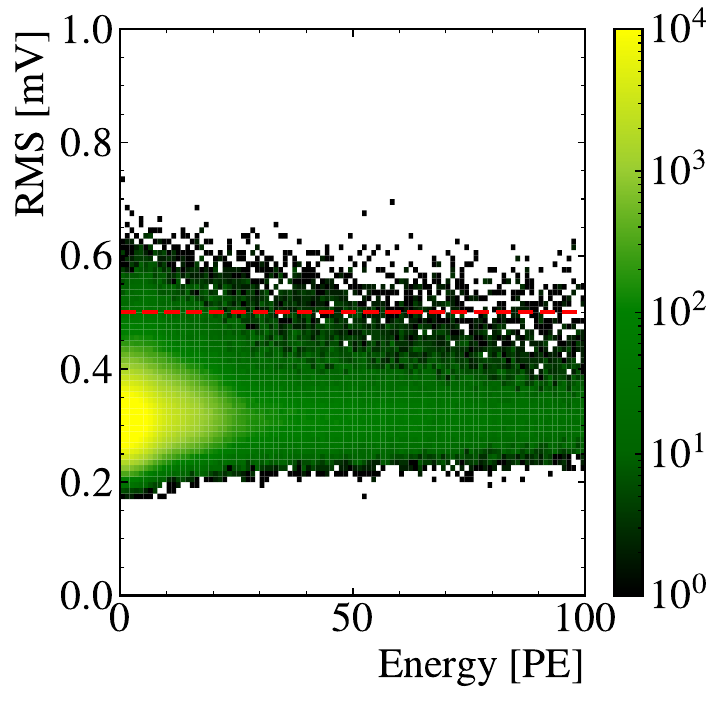}
         \includegraphics[height = 6cm]{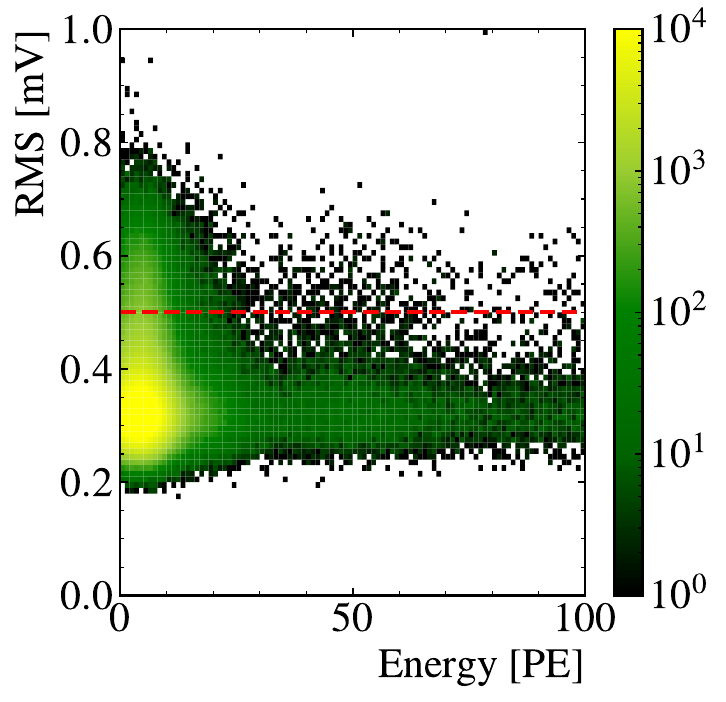}
        
         \includegraphics[height = 6cm]{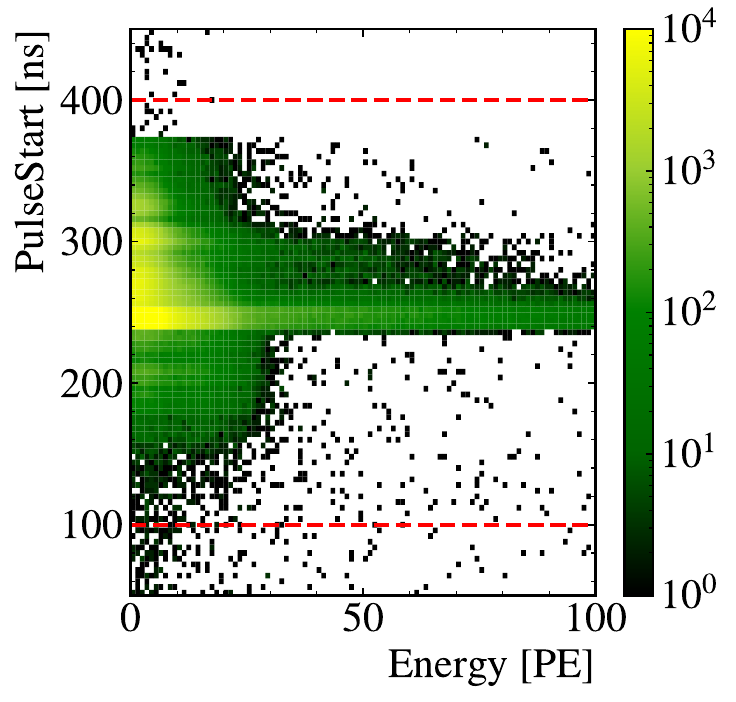}
         \includegraphics[height = 6cm]{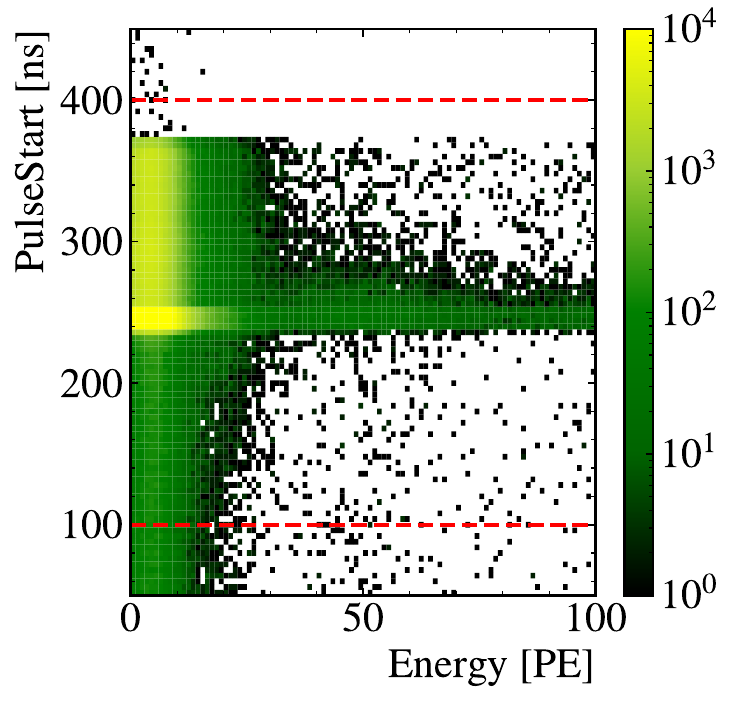}
         
    \caption{Selection cuts made on the dark and \Ba{} datasets.  (Top) RMS ($\leq$ 0.5~mV) and (Bottom) Pulse start locations ($100~\rm{ns}\leq \text{x}\leq400~\rm{ns}$). (Left) background and (Right) signal.}
    \label{fig:TrainingCuts}
\end{figure}

The selection criteria for the training population are as follows:  cosmic muon background is removed if a coincident signal is found in the  scintillation panels: RF events are removed by requiring $\mathcal{L}_{\rm{RF}}$$\geq$ -0.25, and events are required to have 5~--~30 PEs, which corresponds to the sub-3~\kevee region where PMT background has been observed to dominate~\cite{Coarasa_2022}. For both datasets, we apply additional criteria by requiring they have a baseline RMS ~$\leq$~0.5~mV and 100~$\leq$~pulse start~$\leq$~400 ns (Fig.~\ref{fig:TrainingCuts}). Large RMS scores are consistent with events that have peaks and/or structure in the baseline estimation window (i.e.,the first 100~ns of an event). These selection criteria are applied on the training and testing samples. We ensure that within each energy bin there is an equal number of background and signal events by randomly sampling the two populations. After these selection requirements, we are left with 20 000 events for each PMT with a 50-50 split between background and signal datasets. This is then split into training (60~$\%$) and testing (40~$\%$) samples. We quantify the signal to background separation power of each variable by evaluating~\cite{hoecker2009tmva}:
\begin{equation}\label{eqn:separation}
    \langle S^{2}\rangle \ =\frac{1}{2}\int\frac{(\hat{y}_{S}(y) - \hat{y}_{B}(y))^{2}}{\hat{y}_{S}(y) + \hat{y}_{B}(y)} dy.
\end{equation}
Prior to training, this calculates the degree of overlap between the signal and background PDF ($\hat y_{S}$ and $\hat y_{B}$) for each variable in the combined training and testing samples (higher score implies greater separation). The results are shown in Table~\ref{tab:separationvariables}. We find that the top six parameters have similar separation powers, with slightly different ordering between the three models

\begin{table}[htb]
\centering
\caption{Variable separation, defined in Eq.~\ref{eqn:separation}, between signal ($\gamma$) and background (dark) prior to training for events in the 5~--~30 PE region. The combined dataset is a 50-50 mix of events from both BC0174 and BC0175.}
\label{tab:separationvariables}
    \renewcommand{\arraystretch}{1.1} 
\begin{tabular}{|l|ccc|}
\hline
& \textbf{Combined} [$\%$]& \textbf{BC0174} [$\%$] &\textbf{BC0175} [$\%$]\\
 \hhline{|=|===|}\textbf{CAP100}& 12.9& 12.7&13.7\\
\textbf{X2}& 12.5& 12.4&13.2\\
\textbf{$\mathcal{L}_{\rm{RF}}$}& 12.4& 12.7&12.2\\
\textbf{X1}& 12.0& 11.9&12.7\\
 \textbf{N$_{p}$}& 11.9& 10.4&14.1\\
\textbf{log($\mu_{p}$)} & 10.8& 10.5&11.3\\
\textbf{Skew}& 6.9& 6.9&7.1\\
 \textbf{Kurtosis}& 4.9& 5.1&5.0\\
\textbf{CAP500}& 4.6& 4.8&4.7\\
\textbf{CAP50}& 3.4& 3.6&3.5\\
\textbf{CAP10} & 2.5& 2.8&2.5\\
\hline
\end{tabular}

\end{table}

\subsection{Single PMT model}
A total of three training datasets are used \DSA{}, \DSB{} and a sample that includes data from both \DSA{} and \DSB{} denoted the ``Combined" dataset. This results in the production of 3 models named \MDA{}, \MDB{}, and \MDC{}, each trained using the corresponding dataset. The models are developed using XGBoost~\cite{Chen_2016} through the binary logistic model with the aim of developing a method to separate PMT noise from real events and to understand the dependence of the response model on a PMT basis. The following parameters are set in the model: the learning rate is 0.5, the maximum depth is 3, and there are 200 estimators. The gain  ranking, which measures the improvement in accuracy achieved by introducing the feature to the branch of the tree is listed in Table~\ref{tab:featureimportance}, which are evaluated using the relative contribution of a feature to the model. A difference in the feature importance between the two PMTs is observed. The BC0175 PMT training finds that the most important input is $\mathcal{L}_{\rm{RF}}$, suggesting that there may be more RF events remaining in the BC0175 dataset. In general the BC0174 model, and the combined model have similar importance ranking. The variable CAP100 provides the highest degree of separation, as is the case for variable separation, followed by $\mathcal{L}_{\rm{RF}}$ and N$_{p}$.

\begin{table}[htb]
    \centering
    \caption{Feature importance ranked by gain as identified by each of the three trained models, \MDA{}, \MDB{} and \MDC{}.  The table is ordered based on the Combined model. \\}
    \label{tab:featureimportance}
    \renewcommand{\arraystretch}{1.1} 
    \begin{tabular}{|l|ccc|}
    \hline
    {} &  \textbf{Combined} [$\%$]&  \textbf{BC0174} [$\%$]&  \textbf{BC0175} [$\%$]\\
    \hhline{|=|===|}
    \textbf{CAP100}& 33.6& 35.9&27.3\\
    \textbf{$\mathcal{L}_{\rm{RF}}$}& 30.2& 29.7&32.1\\
    \textbf{N$_{p}$}&        12.2&        11.7&        14.7\\
    \textbf{Skew}&        6.9&        5.0&        6.0\\
    \textbf{X2}&         3.7&         3.5&        7.6\\
    \textbf{CAP50}&        3.6&         3.3&        3.1\\
    \textbf{CAP10}&         2.3&         2.1&         2.2\\
    \textbf{X1}&         2.2&         2.2&         2.4\\
    \textbf{Kurtosis}&         2.0&         2.6&         1.5\\
    \textbf{log($\mu_{p}$)} &         1.8&         2.1&         1.4\\
    \textbf{CAP500}&         1.4&         1.8&         1.5\\
    \hline
    
    \end{tabular}
\end{table}

\begin{figure}[htb]
    \centering
    \includegraphics[width=6 cm]{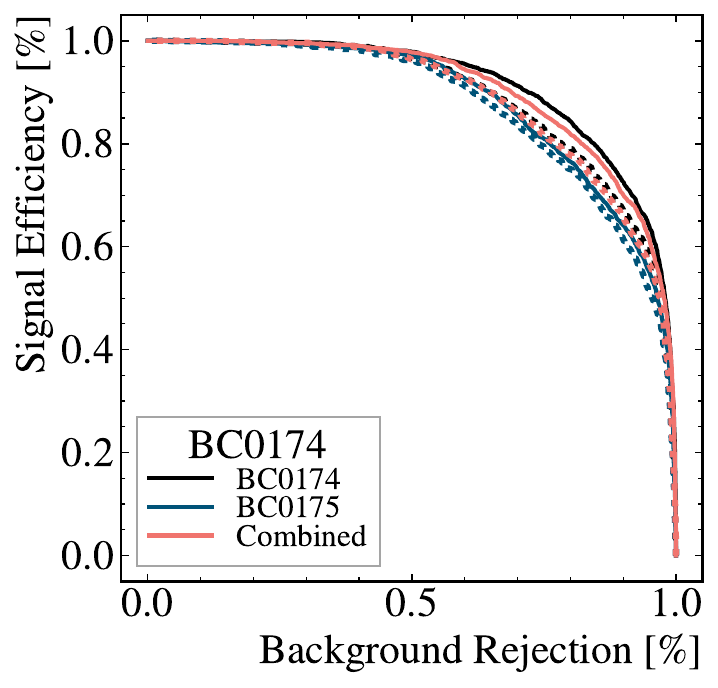}
    \includegraphics[width=6 cm]{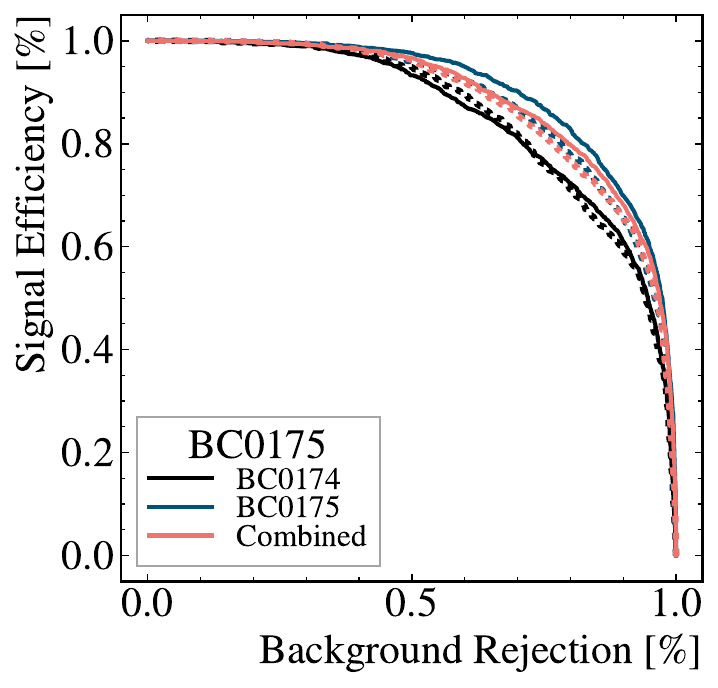}
    
    \hfill
    \caption{The receiver operating curves (ROC) displays the signal efficiency to background rejection for events in the 5~--~30 PE regions. (Left) Results from only BC0174 data and (Right) from only BC0175 data. The different colours represente the different training models used. The different line styles correspond to results from the training (solid lines) and the testing samples (dashed lines).}  
     
    \label{fig:ROC}
\end{figure}

We test each training model with each of the datasets from the BC0174 and BC0175 PMTs. This is to provide insight into PMT dependence in the training and whether it is sufficient to combine training data sets. The receiver operator curve (ROC) for the 5~--~30 PE region is shown in Fig.~\ref{fig:ROC} which displays the signal efficiency of the model as a function of the background rejection.  
\begin{align}\label{eq:eff_definitions}
\text{Signal Efficiency } &= \frac{N_{\text{Correctly classified signal events} }}{N_{\text{Signal Events}}}, \\
\text{Background Rejection} &= 1-\frac{N_{\text{Incorrectly classified background events}}}{N_{\text{Background events}}}
\end{align}
Performance of the models on data taken from BC0174 and BC0175 showed a significant difference in the signal efficiency and background rejection. In both cases the combined model performed similarly with the alternate model (i.e. using model BC0174, to predict events from the BC0175 dataset) performed worse than the other two models. The combined model performed similarly in both situations.

\subsection{Validation}

\begin{figure}[htb]
    \centering
    \captionsetup[subfigure]{justification=centering}
        \includegraphics[height = 6cm]{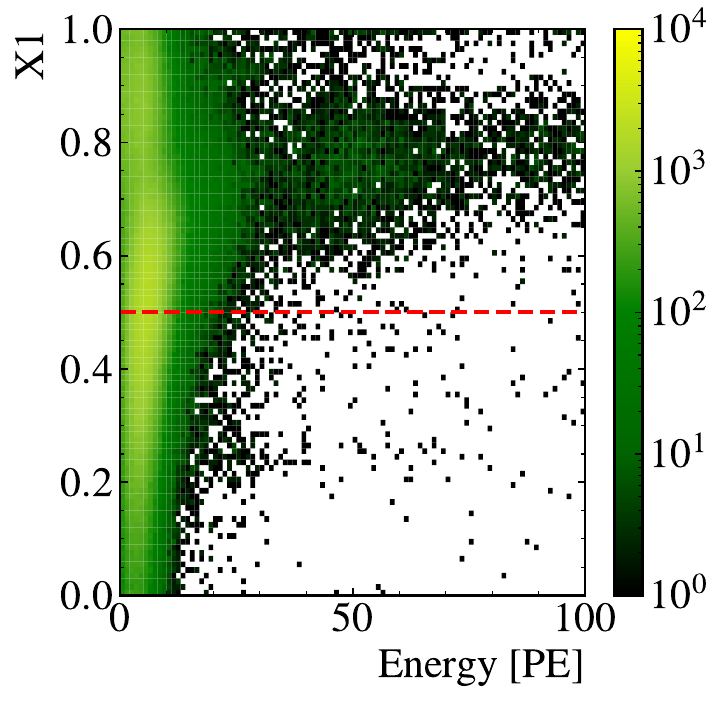}
        \includegraphics[height = 6cm]{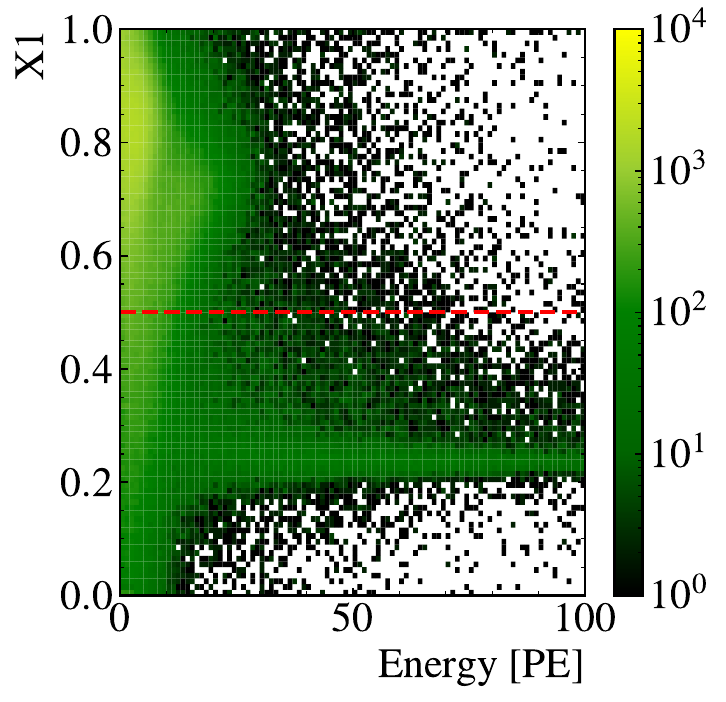}
    \hfill
        \includegraphics[height = 6cm]{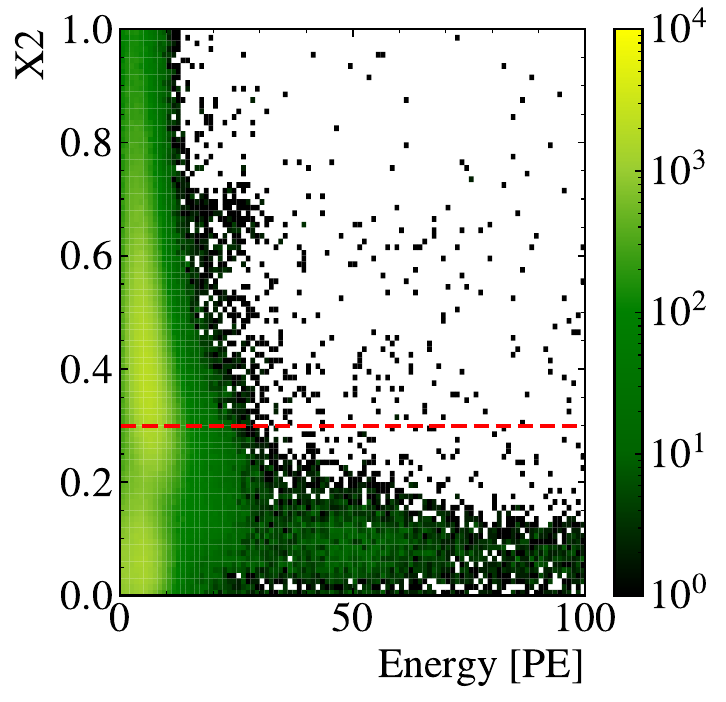}
        \includegraphics[height = 6cm]{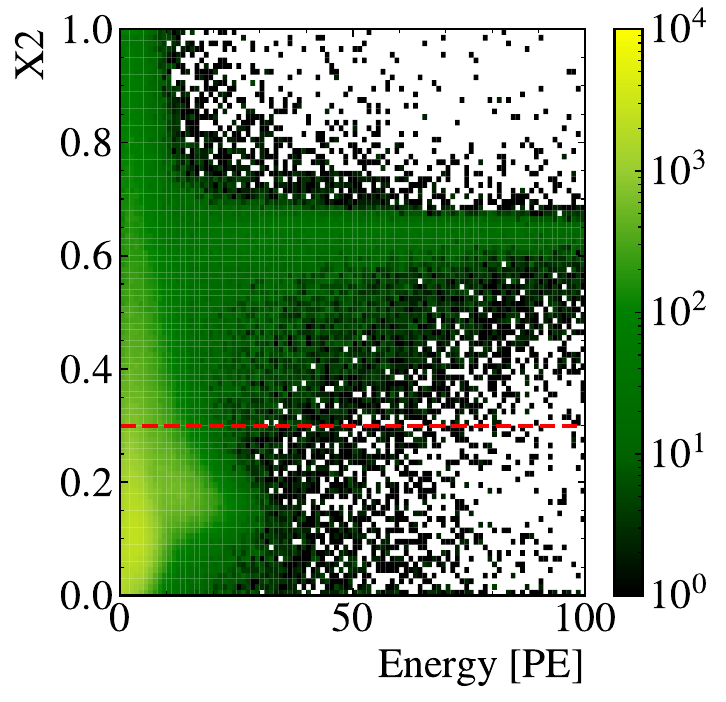}
    \caption{Distribution of X1 (Top) and X2 (Bottom) vs energy in photoelectrons for BC0175. Selection cuts are applied on both the signal (left) and background (right) datasets. The shaded red region shows the excluded region.   
    }
    \label{fig:DamaCuts}
\end{figure}

\begin{figure}[htb]
    \centering
    \includegraphics[height=6 cm]{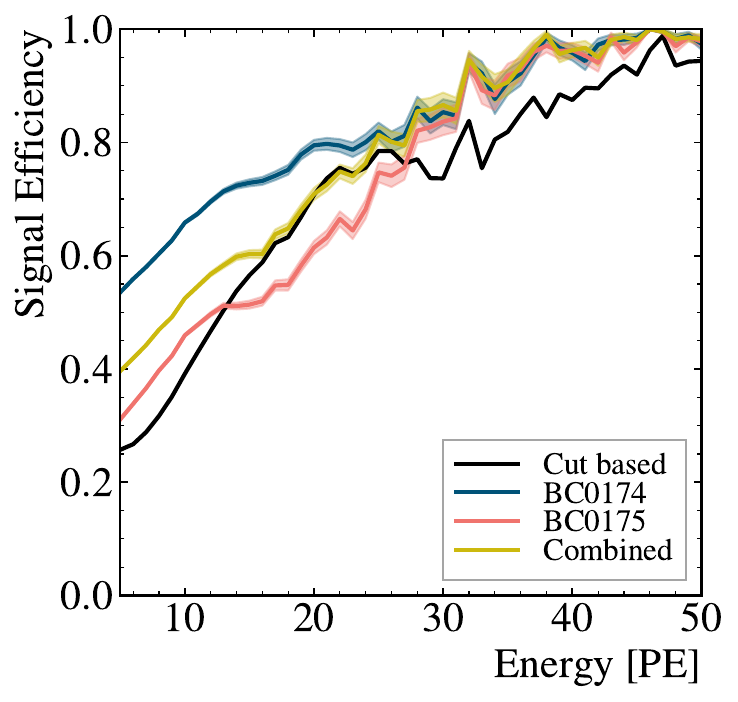}
    \includegraphics[height=6 cm]{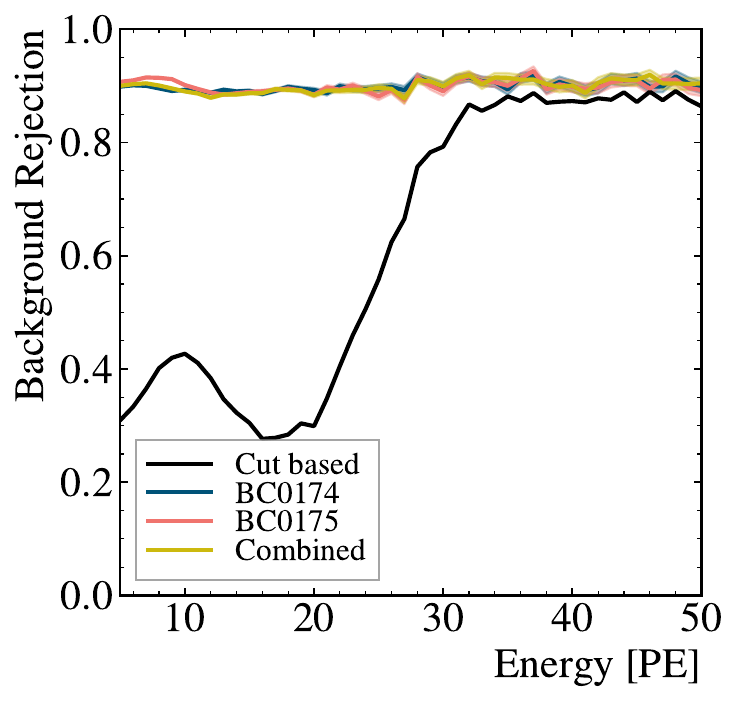}
    \caption{(Left) The signal efficiency, with the dashed line indicating the 10 PE limit and (Right) background rejection in 1 PE bins. The thresholds used to cut was based on the combined classifier. The errors are calculated using a binomial error.  
    }  
    \label{fig:SigEffFinal}
\end{figure}

We compare the performance of the BDT against a cut based on X1 and X2, to remove electronic noise from the signal~\cite{DAMA/LIBRA_apparatus}. We characterise performance by evaluating the signal efficiency at  90~$\%$ background rejection for the full dataset excluding muon events, for our BDT model in each 1 PE bin. We additionally created a cut based baseline using the X1 and X2 parameters used by DAMA/LIBRA, these distributions are shown (for BC0175) in Fig.~\ref{fig:DamaCuts} for data from both BC0174 and BC0175 (combined). The peak in the background samples appears at X1 = 0.2 and X2 = 0.7, therefore, we apply selections that require X1$>$ 0.5 and X2 $<$ 0.3, providing a background rejection of approximately 90~$\%$ above 30 PE.  The resulting signal efficiency and background rejection are shown Fig.~\ref{fig:SigEffFinal}. In all three models we obtain a signal efficiency better than our baseline cut-based method. We observe a significant performance gap between the best- and worst-performing models. This study shows that, at the targeted 10~PE threshold, we can achieve up to a 20~\% increase in signal efficiency over the cut-based method while also improving background rejection at lower PEs. The large variation in BDT performance suggests that certain individual PMT characteristics may influence overall signal-background discrimination capabilities.

\subsection{PMT coincidence}

\begin{figure}[htb]
    \centering
    \captionsetup[subfigure]{justification=centering}
    \begin{subfigure}[b]{0.9\columnwidth}
         \centering
        \includegraphics[height=5 cm]{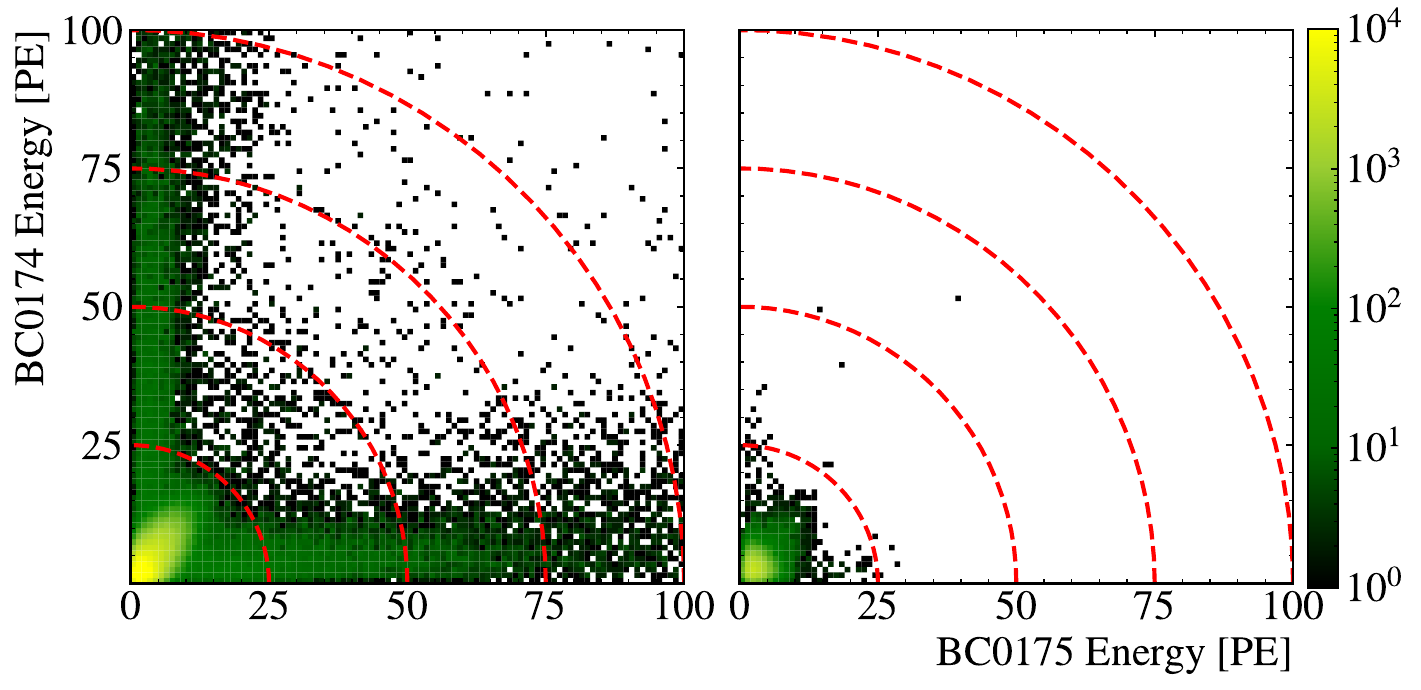}
         \caption{Noise classified}
         \label{fig:noiseCl}
    \end{subfigure}
    \hfill
    \begin{subfigure}[b]{0.9\columnwidth}
         \centering        
         \includegraphics[height=5 cm]{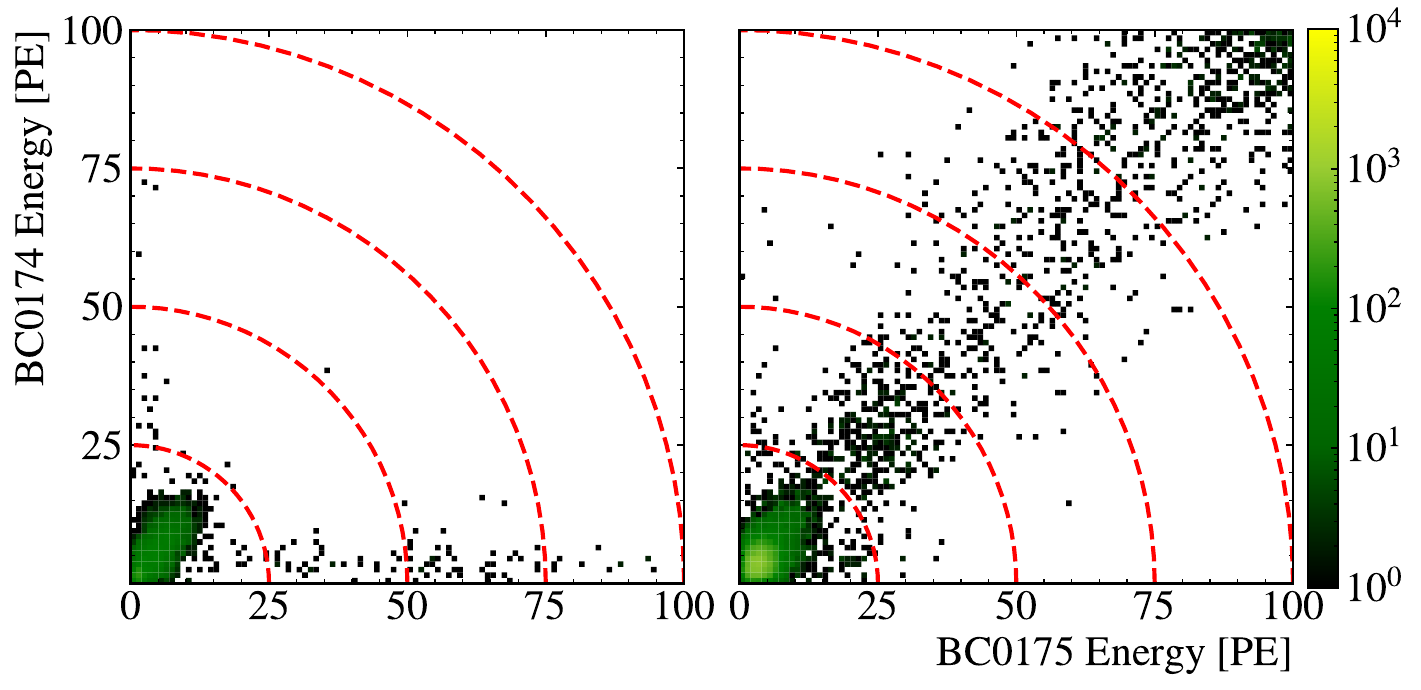}
         \caption{Signal classified}
         \label{fig:signalCl}
    \end{subfigure}
    \caption{ The classified signal and background based on the \MDC{} classifier. Dark (Left) and Signal (Right) data. The contours are shown at 20 PE intervals. The signal classification follows a logical "OR" hence only a single PMT has to classify the event as a signal. No coincidence information was used to obtain these results.}
    \label{fig:2dclassified}
\end{figure}

We expect in PMT pairs, scintillation light originating at the centre of the crystal will induce similar sized pulses in each PMT. Figure~\ref{fig:2dclassified} shows the events classified as signal (if at least one PMT classified the event as a signal candidate) and background according to our combined model. In the noise-classified background and signal events a majority have a low number of PEs. Amongst the noise data set the majority of events are highly asymmetric in energy deposition. In the signal data set we also observe a portion distributed asymmetrically.  

Most events classified as noise in the signal dataset fall below 20~PE, with a tail along each axis corresponding to asymmetric photon depositions. This could result from dark rate contributions at low PE counts or real scintillation events occurring near or at the window of a single PMT. We observe that the model misclassifies more frequently at low PE counts. The classifier effectively identifies signal events above 20 PE. Other collaborations have explored methods to measure asymmetry in photon deposition between two PMTs~\cite{Coarasa_2022}. 

\subsection{Paired PMT model}
As a final comparison, a na\"ive approach is taken by combining variables from both PMTs, making use of parameters from BC0174 and BC0175 as a paired PMT model. As information from both PMTs is utilised, an additional parameter can be introduced, Asymmetry$_{\text{NPE}}$. This quantifies the degree of asymmetry in PE observed between the two PMTs, and is defined as
\begin{equation}\label{eq:asymmetry}
    \text{Asymmetry}_{\text{NPE}} = \frac{\text{NPE}_{A} - \text{NPE}_{B}}{\text{NPE}_{A} + \text{NPE}_{B}}.
\end{equation}
The same selection criteria are applied for our training data set as in the single PMT model (see Section.~\ref{sec:PMT_bdt_selection}), where both the BC0174 and BC0175 pass the selection criteria for RMS, pulse start and coincidence. The RF requirement is set to $\mathcal{L}_{\rm{RF}}$$\geq -0.3$, providing 7\,645 events for training. The top 10 parameters for this model are shown in Table.~\ref{tab:pair_feature_of_importance}, as well as the performance of the asymmetry$_{\text{NPE}}$ parameter.

\begin{table}[htb]
    \centering
    \caption{The gain feature importance for the paired BDT model. The top 10 parameters are shown along with the asymmetry parameter performance}
    \label{tab:pair_feature_of_importance}
    \renewcommand{\arraystretch}{1.1} 
    \begin{tabular}{|l|c|c|}
        \hline
        Feature & PMT &Value [$\%$] \\
        \hhline{|=|=|=|}
        $\mathcal{L}_{RF}$ & BC0174 & 19.11 \\
        CAP100 & BC0175 & 13.80 \\
        $\mathcal{L}_{RF}$ & BC0175 & 12.87 \\
        N$_{p}$ & BC0175 & 9.57 \\
        N$_{p}$ & BC0174 & 9.15 \\
        CAP100 & BC0174 & 9.15 \\
        X2 & BC0174 & 4.00 \\
        X2 & BC0175 & 3.22 \\
        Skew & BC0175 & 2.71 \\
        Skew & BC0174 & 2.29 \\
        $\cdots$&$\cdots$&$\cdots$\\
        Asymmetry$_{\text{NPE}}$& Both &1.10 \\
        \hline
    \end{tabular}
\end{table}

\begin{figure}[htb]
    \centering
    \includegraphics[height=6cm]{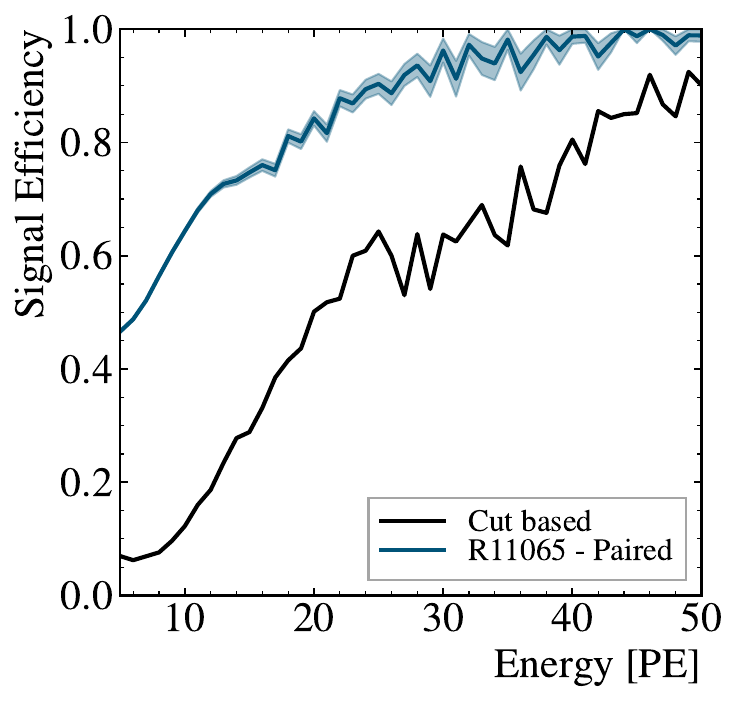}
    \includegraphics[height=6cm]{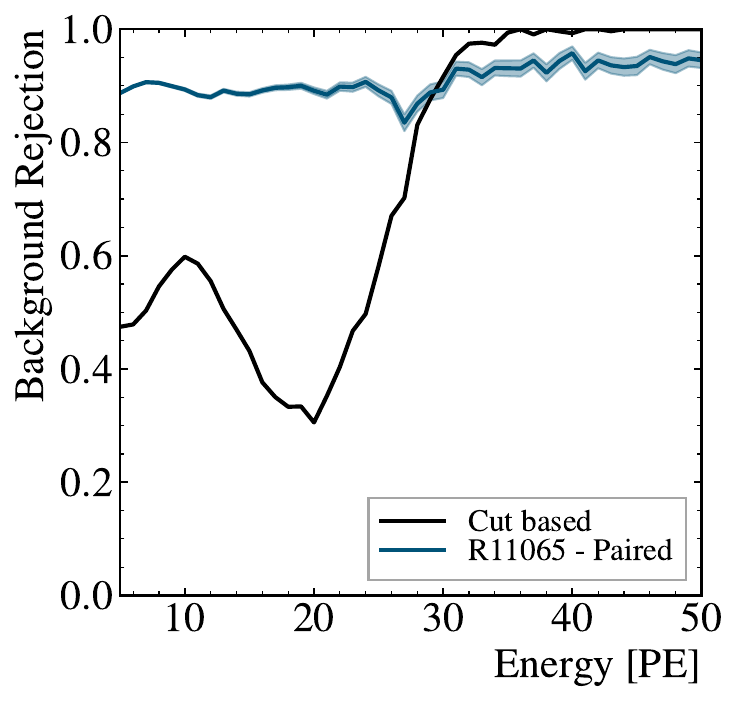}
    \caption{(Left) Signal efficiency and (Right) background rejection performance for the paired PMT model. The errors are calculated using a binomial error. }
    \label{fig:PairedSignalEfficiency}
\end{figure}

The results show that the same parameter pairs (regardless of PMT) has the greatest importance for the model, i.e. $\mathcal{L}_{RF}$ for BC0174 and BC0175 are both ranked in the top three. Similar to the single PMT model, we also determine the signal efficiency and background rejection for the cut based model by requiring that both PMTs pass X1$\geq$~0.6 and X2$\leq$0.3. The result of the model, as well as these selection criteria, are shown in Fig.~\ref{fig:PairedSignalEfficiency}.

Figure~\ref{fig:PairedSignalEfficiency} shows a significant increase in signal efficiency from the cut-based model below 50 PE. The paired model also shows better performance than the single PMT approach shown in Fig.~\ref{fig:SigEffFinal}, compared to the combined model (as a conservative comparison) we have a 10$\%$ improvement at 5 PE and 20$\%$ improvement at 10 PE. The paired and single PMT performances are the same above 35 PE ($\approx 3$ \kevee), which matches the point that other experiments \cite{Coarasa_2022} have observed their collected data depart from simulation and benefit most from BDTs. The results have shown that application of these BDT models will allow us to reduce our detector thresholds down to 5 PE with 90$\%$ background rejection.

\section{Waveform Simulation}
The results from pre-calibration and characterisation described in this paper are input to a waveform simulation tool developed for use in data analysis, denoted DOOM (Digitisation of Optical Monte Carlo simulation). These fully simulated waveforms are used to study the impact of intrinsic noise and the quantised nature of photon counting on energy resolution, background rejection, and event reconstruction. The algorithm takes the output of optical simulations from the Geant4-based SABRE simulation~\cite{SABRE_MC} and produces synthetic events that can be analysed as real data using the SABRE South processing framework. The code simulates the physical process of photoelectron production and the resulting avalanche for each photon that reaches the photocathode, and uses the SPE charge model detailed in Equation~\ref{eq:spe_q_model}. 

The photoelectron charge is introduced as a delta function into a time-binned waveform with a finer time binning than the final output. The time-binned charge distribution is convolved with a SPE probability distribution function to produce an `ideal' waveform. This is rebinned to the nominal sampling rate of the digitiser before the baseline waveform, with appropriate variation, is added. The final waveform is then passed through a simulated trigger. Thermionic dark events are simulated in DOOM by overlaying additional photoelectrons at random times to the initial simulation input. 

\begin{figure}[htb]
    \centering
    \captionsetup[subfigure]{justification=centering}
         \centering
        \includegraphics[height=5.5cm]{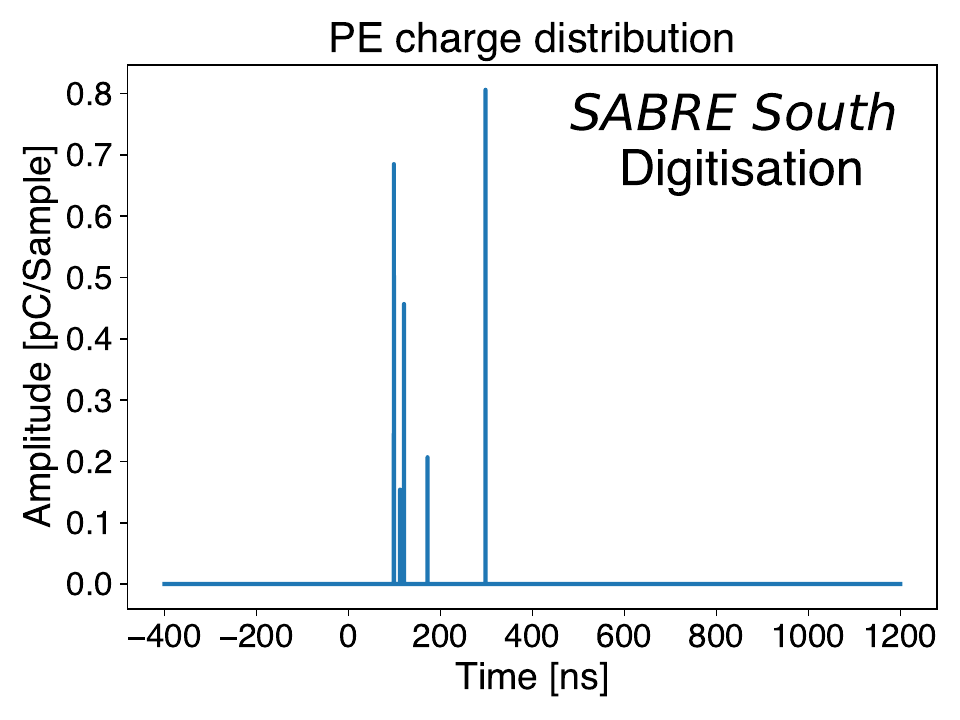}
    \hfill
         \centering        
         \includegraphics[height=5.5cm]{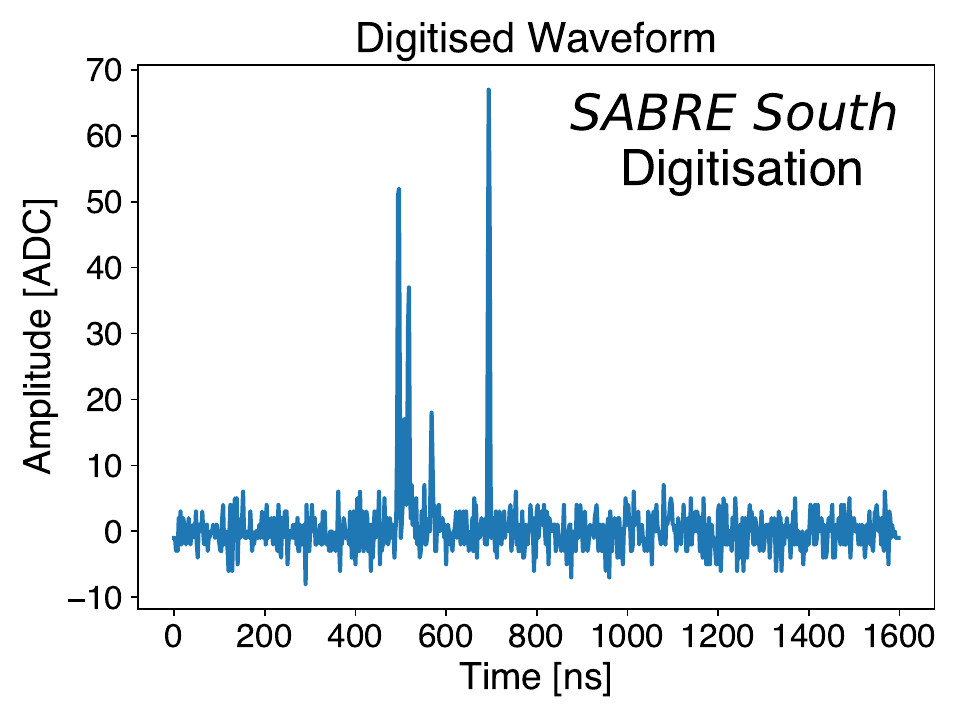}
    \caption{Example digitised waveforms of 1~keV gamma ray in SABRE \nai{} detector module. (left) Initial distribution of PE with charges drawn from PE charge model (right) The resulting digitised waveform after the digitisation process including the resolution and baseline modelling.}
    \label{fig:DOOMWaveform}
\end{figure}

The number of added thermionic electrons is determined using a Poisson distribution determined by the measured rate of thermionic emissions from fits to the SPE equivalent component of the dark-event charge spectra. For the R11065 photomultiplier, the majority of the dark events detected in Section~\ref{sec:dr} are thermionic. An example of the resulting waveforms is shown in Fig.~\ref{fig:DOOMWaveform}.

To accurately simulate the response of the detector, the performance characteristics of each photomultiplier in the SABRE South experiment are implemented. DOOM takes as input the quantum efficiency, the transit-time mean and standard deviation, the thermionic dark rate at the SABRE operating temperature in SUPL, the SPE charge model at the operating voltage, and the SPE pulse shape. 

The baseline variation is modelled as a Gaussian using a mean and variation determined from the baseline samples of the above analysis.  This tool has been validated against above ground measurements with NaI(Tl) and a variety of sources.

The primary use of this tool is to overcome the challenges of producing high purity low energy scintillation signal datasets, which can be used in the development of noise rejection tools optimised for the 1~\kevee{} energy threshold. These methods have been applied in other experiments~\cite{XENONCollaboration:2023dar, Choi:2024ziz}. Additionally, it can be used to quantify the trigger and selection efficiency at 1 \kevee{}, as well as the impact of dark events on the timing and position reconstruction of low-photon number scintillation events (order 10 photo-electrons) in both NaI(Tl) detectors and in the SABRE liquid scintillator veto.

\section{Conclusion}

This paper has outlined the pre-calibration methods for the SABRE South PMTs. This was demonstrated by looking at two of the R11065 PMTs that will be coupled to the NaI(Tl) crystals, displaying the capability to measure the characteristics of the PMTs (gain, dark rate, and timing), as well as a method for long-term gain tracking within the vessel by use of dark events, which has the potential to be conducted live. Similar procedures are used for the pre-calibration procedure of the SABRE South R5912 veto PMTs which will be reported in future work. Exploration of noise discrimination methods using data from single PMTs, showed the ability to improve on cut-based methods employed by DAMA/LIBRA below 30 PE. The model has shown difficulties in separating highly asymmetric energy deposition and indicated the possibility that PMT characteristics such as changes in the dark rate and gain may play a larger role in the reliability of these models. Investigation into the cause of these effects is beyond the current scope of this work. We additionally compared the performance of a na\"ive pairwise approach to our single PMT BDT models, and found that we can expect a $20~\%$ signal efficiency improvement at 10 \kevee over our single PMT models. We further briefly outline the development of a waveform simulation tool, which will be used for the development of future machine learning models and analysis for the SABRE South detector, taking in the measured PMT characteristics to produce realistic digitised waveforms for training.

\newpage
\section*{Acknowledgements}
The SABRE South program is supported by the Australian Government through the Australian Research Council (Grants: CE200100008, LE190100196, LE170100162, LE160100080, DP190103123, DP170101675, LP150100705). This research was partially supported by Australian Government Research Training Program Scholarships, and Melbourne Research Scholarships. This research was supported by The University of Melbourne's Research Computing Services and the Petascale Campus Initiative.
Technical Support from the ANU Workshop (Daniel Tempra, Thomas Tunningley).

\bibliographystyle{JHEP}
\bibliography{references}
 
\end{document}